\documentclass[11pt, draftclsnofoot, onecolumn]{IEEEtran}
\usepackage{cite}
\usepackage{url}
\usepackage{color}
\usepackage{overpic}
\usepackage{bm}

\usepackage[dvipsnames,svgnames]{xcolor}
\usepackage{graphicx}
\usepackage{psfrag}
\DeclareGraphicsExtensions{.xps}
\usepackage{epstopdf} 
\usepackage{enumerate}
\usepackage{subcaption}
\usepackage{dirtytalk}
\usepackage[font=small,labelfont=bf]{caption}
\usepackage[cmex10]{amsmath}
\usepackage{amssymb}
\usepackage{amsthm}

\usepackage[utf8]{inputenc}
\usepackage[T1]{fontenc}

\newcommand{\vect}[1]{\boldsymbol{#1}}
\def\pb{\textrm{pb}}
\def\pos{\vect{p}}
\def\beam{Q}
\def\beamind{i}
\def\posUE{\vect{p}}
\def\sinc{\mathrm{sinc}}
\def\Htran{\mbox{\tiny $\mathrm{H}$}}
\def\Ttran{\mbox{\tiny $\mathrm{T}$}}
\def\CN{\mathcal{N}_{\mathbb{C}}} 
\def\imagunit{\mathsf{j}} 
\newcommand{\argmax}[1]{{\underset{{#1}}{\mathrm{arg\,max}}}}

\begin{document}

\IEEEoverridecommandlockouts

\title{Reconfigurable Intelligent Surfaces: \\ \LARGE{A Signal Processing Perspective With Wireless Applications}}
\author{
\IEEEauthorblockN{Emil Bj{\"o}rnson, Henk Wymeersch, Bho Matthiesen, Petar Popovski, Luca Sanguinetti, and Elisabeth de Carvalho \thanks{E.~Bj\"ornson is with KTH Royal Institute of Technology, 100 44 Stockholm, Sweden (emilbjo@kth.se).
H.~Wymeersch is with Chalmers University of Technology, 412 96, Sweden (henkw@chalmers.se). 
B.~Matthiesen is with University of Bremen, 28359 Bremen, Germany (matthiesen@uni-bremen.de).
P.~Popovski and E.~de Carvalho are with Aalborg University, 9220 Aalborg, Denmark (\{petarp,edc\}@es.aau.dk).
L.~Sanguinetti is with University of Pisa, 56125 Pisa, Italy (luca.sanguinetti@unipi.it).
}}
}
\maketitle

\vspace{-6mm}

Antenna array technology enables directional transmission and reception of wireless signals, for communications, localization, and sensing purposes. The signal processing algorithms that underpin this technology began to be developed several decades ago \cite{Veen1998a}, but it is first with the ongoing deployment of the fifth-generation (5G) wireless mobile networks that it becomes a mainstream technology \cite{Bjornson2019d}. The number of antenna elements in the arrays of the 5G base stations and user devices can be measured at the order of 100 and 10, respectively. As the networks shift towards using higher frequency bands, more antennas fit into a given aperture. For communication purposes, the arrays are used to form beams in desired directions to improve the signal-to-noise ratio (SNR), multiplex data signals in the spatial domain (to one or multiple devices), and suppress interference by spatial filtering \cite{Bjornson2019d}. For localization purposes, these arrays are used to maintain the SNR when operating over wider bandwidths, for angle-of-arrival estimation, and to separate multiple sources and scatterers \cite{Witrisal2016}. 
The practical use of these features requires that each antenna array is equipped with well-designed signal processing algorithms.

\begin{figure}[t!]
        \centering
        \begin{subfigure}[b]{\columnwidth} \centering  
	\begin{overpic}[width=.9\columnwidth,tics=10]{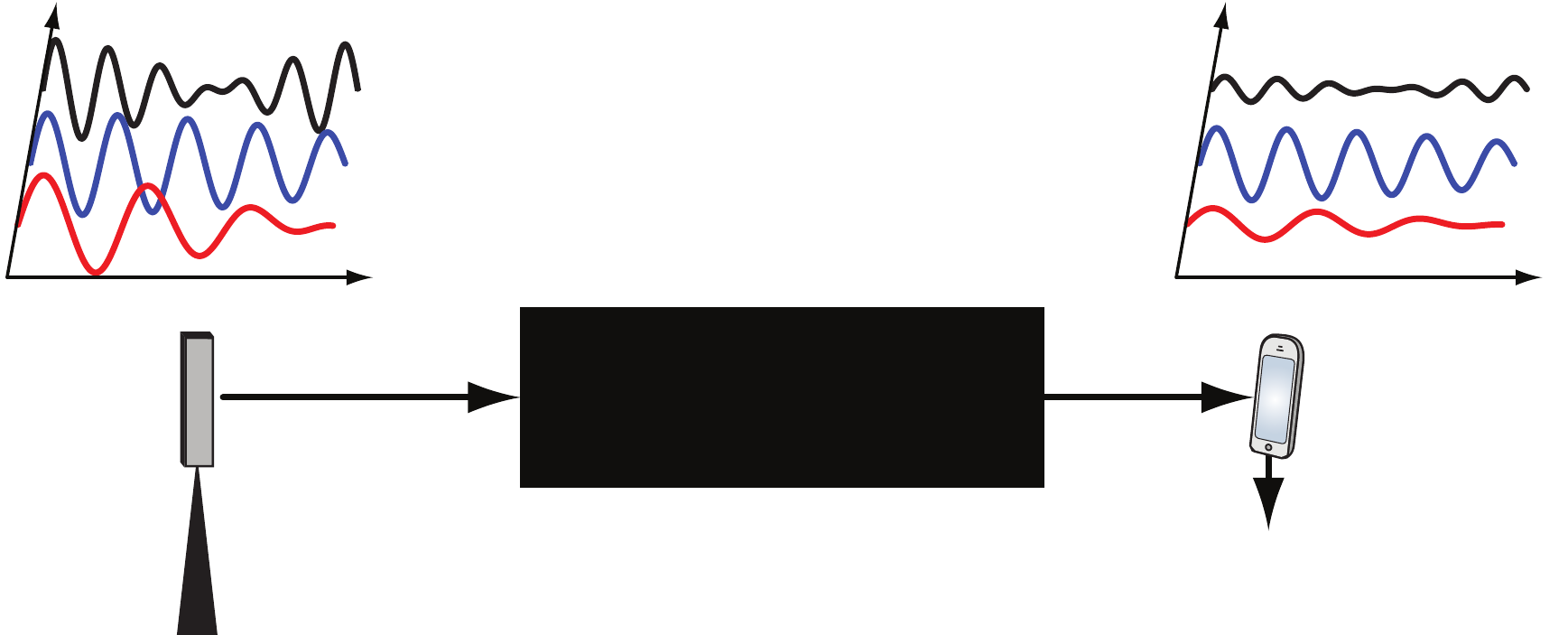}
		\put(20,24){\footnotesize Time}
		\put(1,41){\footnotesize Frequency}
		\put(95,24){\footnotesize Time}
		\put(75.5,41){\footnotesize Frequency}
  \put (-1,4) {Transmitter}
  \put (84,16) {Receiver}
  \put(35.5,16.6) {\color{white}Uncontrollable propagation}
  \put(36,12.6) {\color{white}Impulse response: $h_{d,\pb}(t)$}
  \put(14.5,17) {$x_{\pb}(t)$}
  \put(77,4) {$y_{\pb}(t)= (h_{d,\pb} *x_{\pb})(t)$}
\end{overpic} 
                \caption{Conventional wireless system where the channel propagation is uncontrollable.} 
                \label{fig:basic_example_a}
        \end{subfigure} 
        \begin{subfigure}[b]{\columnwidth} \centering 
	\begin{overpic}[width=.9\columnwidth,tics=10]{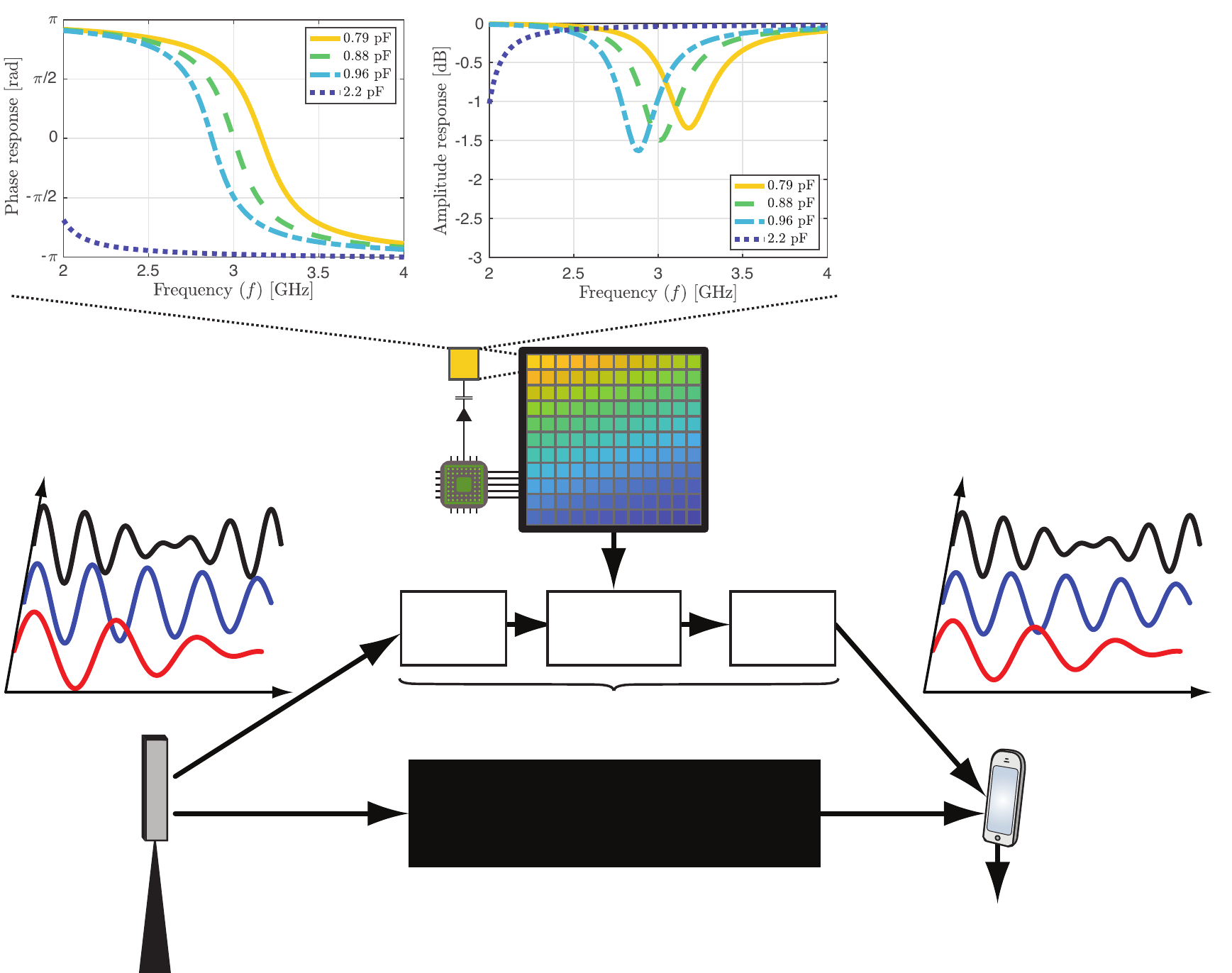}
  \put (-1,4) {Transmitter}
  \put (84,14) {Receiver}
  \put(35.5,14.6) {\color{white}Uncontrollable propagation}
  \put(36,10.6) {\color{white}Impulse response: $h_{d,\pb}(t)$}
  \put(15.5,15) {$x_{\pb}(t)$}
  \put(57,3) {$y_{\pb}(t)= \sum\limits_{n=1}^{N} (h_{\pb; \theta_n} *x_{\pb})(t) + (h_{d,\pb} *x_{\pb})(t)$}
		\put(25.5,49){\footnotesize Element $n$}
		\put(30,45){\footnotesize Switch}
		\put(25.5,39){\footnotesize Controller}
		\put(20,24){\footnotesize Time}
		\put(1,41){\footnotesize Frequency}
		\put(95,24){\footnotesize Time}
		\put(75.5,41){\footnotesize Frequency}
  \put(69,69) {Frequency-domain}
  \put(69,66) {representation of $\vartheta_{n,\pb; \theta_n}\!(t)$}
 \put (58.5,44) {RIS with}
 \put (58.5,41) {$N$ elements}
 \put (32,34) {\footnotesize Channel to }
 \put (33,32) {\footnotesize  element}
  \put (57.5,34) {\footnotesize Channel from}
  \put (60,32) {\footnotesize  element}
\put(68,29){\footnotesize \rotatebox{-47}{for $n=1,\ldots,N$}}
	\put(33,27){$a_{n,\pb}(t)$}
	\put(44.7,27){$\vartheta_{n,\pb; \theta_n}\!(t)$}
	\put(59.8,27){$b_{n,\pb}(t)$}
	\put(30.5,20.5){$h_{\pb; \theta_n}(t)= (b_{n,\pb} * \vartheta_{n,\pb; \theta_n} * a_{n,\pb})(t)$}
	\put(46,33){$\theta_n$}
\end{overpic}
                \caption{RIS-aided communication system with controllable propagation paths.} 
                \label{fig:basic_example_b}
        \end{subfigure} \vspace{-3mm}
        \caption{The propagation channel in current wireless systems is uncontrollable, as illustrated in (a). When an RIS with $N$ elements is added to the system as in (b), $N$ controllable paths are added to the end-to-end channel. The amplitude and phase of each element can be tuned to improve the signal quality at the receiver.} \vspace{-3mm}
        \label{fig:basic_example}
\end{figure}

The 5G developments enhance the transmitter and receiver functionalities, but the wireless channel propagation remains an uncontrollable system. This is illustrated in Fig.~\ref{fig:basic_example}(a) and its mathematical notation will be introduced later. Transmitted signals with three different frequencies are shown to illustrate the fact that attenuation can vary greatly across frequencies.
Looking beyond 5G, the advent of electromagnetic components that can shape how they interact with wireless signals enables partial control of the propagation. A \emph{reconfigurable intelligent surface (RIS)} is a two-dimensional surface of engineered material whose properties are reconfigurable rather than static~\cite{Tsilipakos2020a}.  As illustrated in Fig.~\ref{fig:basic_example}(b), the surface consists of an array of discrete elements, where each color represents a certain amplitude and phase response curve. A controller and switch determine which curve to utilize, on a per-element or group-of-elements level.
The scattering, absorption, reflection, and diffraction properties of the entire RIS can thereby be changed with time and controlled by software. 
In principle, the surface can be used to synthesize an arbitrarily shaped object of the same size, when it comes to how electromagnetic waves interact with it \cite{Bjornson2020a}.
Fig.~\ref{fig:basic_example}(b) shows how the RIS adds new controllable paths to complement the uncontrollable propagation, each containing a wireless channel to an RIS element, filtering inside the element, and a wireless channel to the receiver.
 These paths can be tuned to improve the channel quality in a variety of ways \cite{Liaskos2018a}.
For example, Fig.~\ref{fig:basic_example}(a) shows how the uncontrollable channel  attenuates some signal frequencies more than others, while 
Fig.~\ref{fig:basic_example}(b) shows how the RIS can be tuned to mitigate this issue. 
An RIS can be utilized to support wireless communications as well as localization, sensing, and wireless power transfer \cite{Renzo2020b,Wymeersch2020b}.

The long-term vision of the RIS technology is to create smart radio environments \cite{Renzo2019a}, where the wireless propagation conditions are co-engineered with the physical-layer signaling, and investigate how to utilize this new capability. The traditional protocol stack consists of seven layers and wireless technology is chiefly focused on the first three layers (physical, link, and network) \cite{Popovski2020a}. 
The conventional design starts at Layer 1, where the physical signals are generated and radiated by the transmitter and then measured and decoded by the receiver. The wireless medium between the transmitter and receiver, called Layer 0, is commonly seen as uncontrollable and decided by ``nature''. The RIS technology changes this situation by extending the protocol design to Layer 0, which can profoundly change wireless systems beyond 5G.

This article provides a tutorial on the fundamental properties of the RIS technology from a signal processing perspective. It is meant as a complement to recent surveys of electromagnetic and hardware aspects \cite{He2019a,Tsilipakos2020a,Renzo2020b}, acoustics \cite{Assouar2018a}, communication theory \cite{Wu2020a}, and localization \cite{Wymeersch2020b}. We will provide the formulas and derivations that are required to understand and analyze RIS-aided systems using signal processing, and exemplify how they can be utilized for improved communication, localization, and sensing. We will also elaborate on the fundamentally new possibilities enabled by Layer 0 engineering and phenomena that remain to be modeled and utilized for improved signal processing design.
The simulation examples can be reproduced using code available at https://github.com/emilbjornson/SPM\_RIS

\section{History and fundamentals}

RIS is an umbrella term that recently appeared in the communication field \cite{Huang2018a}, but the technology has deep roots in the electromagnetic field \cite{Huang2005a,He2019a,Nayeri2015a,Tsilipakos2020a}. There are several decades of research on how to build such surfaces and controlling their properties, and implementation concepts using different materials for different frequencies and use cases. The common feature is that the surface consists of many discrete elements with controllable properties, which are illustrated as colored squares in Fig.~\ref{fig:basic_example}(b).
The elements are passive circuits in the sense that the incoming signals are reradiated after filtering that cannot increase the power. Each element filters the signal by potentially reducing the amplitude, incurring delays, and/or changing the polarization.  
Each element performs this filtering passively based on its local impedance, but the key feature of an RIS is that the impedance can be reconfigured over time by external stimuli.
Fig.~\ref{fig:basic_example}(b) exemplifies how each element is connected by a switch (e.g., a varactor) to a programmable controller that can tune the impedance of the element,
thereby controlling the reflection coefficient that determines the change in amplitude and phase of the reradiated signal \cite{Zhu2013a}. 
The elements are typically sub-wavelength-sized (e.g., a square patch of size $\lambda/5 \times \lambda/5$) to behave as scatterers without a strong intrinsic directivity \cite{Ozdogan2019a}. 
The RIS can then receive signals from any direction from the half-space towards which the RIS elements are facing and tune the pattern of reflection coefficients over the elements to reradiate signals with the desired direction and beam shape. We will later explain the signal processing algorithms that enable this type of operation.

The RIS technology 
appears under different names, such as
software-controlled metasurfaces \cite{Liaskos2018a}, intelligent reflecting surfaces \cite{Wu2019a}, and a few others \cite{Bjornson2019d}. It should be viewed as a general concept for creating smart radio environments where the exact hardware characteristics have been abstracted away. However, it is likely that metasurfaces, where the elements are made of thin layers of metamaterial, will play a major role in practical implementations.
Metamaterials have recently been successfully utilized for commercial antenna design in terrestrial and satellite communications (e.g., by Pivotal Commware 
and Kymeta), as well as radar (e.g., by Echodyne). The RIS technology is different in the sense that the surface is not co-located with the transmitter or  the receiver of the wireless signals but deployed in between, which opens the door for a variety of new use cases as well as new signal processing challenges regarding how to exploit the ability to partially control the channel. The new electromagnetic properties of RIS-aided systems require changes in the established models for discrete signal processing used in communications and localization, and create the need to reexamine the classical system models from first principles to ensure that the technology builds on a solid foundation. The objective of this paper is to provide such a foundation.

\section{End-to-end System Modeling}

The uncontrollable propagation channel in Fig.~\ref{fig:basic_example}(a) is a system that can be analyzed using classical signal processing methods. However, the controllable paths in Fig.~\ref{fig:basic_example}(b) have unusual properties that we will shed light on by providing the connection between the continuous time representations of channels and hardware and the corresponding discrete-time models needed for digital signal processing.

Consider a single-antenna transmitter that sends a wireless passband signal $x_{\pb}(t)$, with time variable  $t \in \mathbb{R}$, to a receiver via an RIS consisting of $N$ scattering elements.
We begin by considering the entire system as uncontrollable; more precisely, it is modeled as linear and time-invariant (LTI) with the real-valued impulse response $h_{\pb}(t)$. It then follows from standard signal processing theory that the output signal $y_{\pb}(t)$ is the convolution between the input and impulse response:
\begin{equation} \label{eq:input-output-passband}
y_{\pb}(t) = (h_{\pb} *x_{\pb})(t) = \int_{-\infty}^{\infty} h_{\pb}(u) x_{\pb}(t-u) du.
\end{equation}
The characterizing feature of an RIS is that its properties can change with time. Hence, this LTI model can only be utilized for the duration of one configuration with a fixed impulse response $h_{\pb}(t)$.
We can distinguish between two RIS regimes: 1) \emph{piecewise constant}, in which $h_{\pb}(t)$ does not change while the signal of interest is non-zero, and 2) \emph{continuously varying}, for which the LTI model in \eqref{eq:input-output-passband} is not valid.
This tutorial focuses on the former category, where the LTI model can be used for the duration of one configuration, but we will briefly describe the second category when discussing mobility effects.

Suppose the transmitted signal is generated from a complex-valued baseband signal $x(t)$ with bandwidth $B/2$ that is modulated to the carrier frequency $f_c$, which satisfies $B\leq 2 f_c$ and usually $B \ll f_c$. For example, a typical scenario in 5G is $f_c=3$ GHz and $B=100$ MHz. The transmitted passband signal will then have bandwidth $B$ and can be expressed as 
\begin{equation} \label{eq:complex-baseband-representation}
x_{\pb}(t) 
= \Re ( \sqrt{2} x(t) e^{\imagunit 2\pi f_c t } ) =  \frac{ x(t) e^{\imagunit 2\pi f_c t } +  x^*(t) e^{-\imagunit 2\pi f_c t }}{\sqrt{2}}
\end{equation}
where $\Re(\cdot)$ outputs the real part of its argument, 
 $ \imagunit=\sqrt{-1}$ is the imaginary unit, and $\sqrt{2}$ keeps the power constant. If we let $\mathcal{F}_c\{\cdot\}$ denote the continuous Fourier transform, the relation in \eqref{eq:complex-baseband-representation} is equivalent to
\begin{equation} \label{eq:complex-baseband-representation-frequency}
X_{\pb}(f) = \frac{X(f-f_c) + X^*(-f-f_c)}{\sqrt{2}} 
\end{equation}
where $X_{\pb}(f) = \mathcal{F}_c \{ x_{\pb}(t) \}$ and $X(f) = \mathcal{F}_c \{ x(t) \}$ are the frequency-domain representations of the passband and baseband signals, respectively. The frequency response $H_{\pb}(f) = \mathcal{F}_c \{ h_{\pb}(t) \}$ describes how the system filters different signal frequencies, generally changing the amplitude and delay differently.

When analyzing passband systems in communication or localization, it is convenient to abstract away the carrier frequency and only consider the baseband signal $x(t)$, which, by definition, has the same power as $x_{\pb}(t)$. We then need to find the baseband counterpart to the input-output relation in \eqref{eq:input-output-passband}.
Many textbooks derive such a relation for the special case of $h_{\pb}(t)$  being a bandlimited passband filter, but we cannot make that assumption since our system describes a  wireless propagation environment that can handle input signals with arbitrary frequency content.
Hence, we will briefly present the so-called complex \emph{pseudo}-baseband representation where the received signal $y(t)$ is defined via $y_{\pb}(t) = \Re ( \sqrt{2} y(t) e^{\imagunit 2\pi f_c t } )$. By taking the Fourier transform of both sides of \eqref{eq:input-output-passband} and utilizing \eqref{eq:complex-baseband-representation-frequency}, we obtain
\begin{equation} \label{eq:input-output-baseband-Fourier1}
Y_{\pb}(f) = H_{\pb}(f)  \frac{X(f-f_c) + X^*(-f-f_c)}{\sqrt{2}} = \frac{ \overbrace{H_{\pb}(f) X(f-f_c)}^{Y(f-f_c)} +  \overbrace{H_{\pb}^*(-f) X^*(-f-f_c)}^{Y^*(-f-f_c)}}{\sqrt{2}} 
\end{equation}
where we use the notation $Y_{\pb}(f) = \mathcal{F}_c \{ y_{\pb}(t) \} = (Y(f-f_c) + Y^*(-f-f_c))/\sqrt{2}$ and $Y(f) = \mathcal{F}_c \{ y(t) \}$. The last equality utilizes the property $H_{\pb}(f) = H_{\pb}^*(-f)$ for real-valued systems. From \eqref{eq:input-output-baseband-Fourier1}, we can identify the Fourier transform of the received baseband signal as
\begin{equation} \label{eq:input-output-baseband-Fourier2}
Y(f-f_c) = H_{\pb}(f) X(f-f_c) \quad \Rightarrow \quad Y(f) = H_{\pb}(f+f_c) X(f).
\end{equation}
Taking the inverse Fourier transform of \eqref{eq:input-output-baseband-Fourier2} yields
\begin{equation} \label{eq:input-output-baseband}
y(t) = (h *x)(t) = \int_{-\infty}^{\infty} h(u) x(t-u) du
\end{equation}
where the impulse response $h(t) = h_{\pb}(t) e^{-\imagunit 2\pi f_c t}$ is the complex pseudo-baseband representation of the system.
The word ``pseudo'' indicates that the downshifted $h(t)$ is not a baseband filter, but the output signal $y(t)$ is anyway baseband since we input the baseband signal $x(t)$. The key benefits of the pseudo-baseband representation is that we can vary the bandwidth $B$ of $x_{\pb}(t)$ without changing the impulse response and that
$h(t)$ represents the true physical system instead of a baseband filtered version of it.

\subsection{Continuous-time system model with RIS elements as reconfigurable filters}

The signal $x_{\pb}(t)$ is the transmitted electromagnetic signal in Fig.~\ref{fig:basic_example}(b) and $y_{\pb}(t)$ is the filtered version that reaches the receiver.
We will now describe how the impulse responses of the RIS paths in Fig.~\ref{fig:basic_example}(b) can be modeled in the pseudo-baseband. For brevity, we only consider the controllable channel via the RIS in this section. We will later enrich the model by including the uncontrollable part, which can describe a
line-of-sight (LOS) path and scattered paths not involving the RIS.
We will characterize the impulse response $h_{\pb;\vect{\theta}}(t)$ of the end-to-end channel filter and we add the subscript
$\vect{\theta}=[\theta_1,\ldots,\theta_N]^{\Ttran}$ to indicate that the impulse response is configured by a set of external control variables $\theta_1,\ldots,\theta_N$ that will be defined below.
For each of the $N$ scattering elements of the RIS, the transmitted signal $x_{\pb}(t)$ will propagate to it over an LTI channel represented by an arbitrary impulse response $a_{n,\pb}(t)$ for element $n = 1, \ldots,N$. If its frequency response has constant amplitude and linear phase over the passband used by the signal, we call it a narrowband channel. We otherwise call it a wideband channel.

When the signal reaches element $n$, it will be filtered inside it and then reradiated. It all happens in the analog domain and we consider a passive operation that can be described by an LTI filter.
The special RIS feature is that the impulse response $\vartheta_{n,\pb; \theta_n}(t)$ is reconfigurable in the sense that it is determined by an external stimulus represented by the variable $\theta_n$. Depending on the RIS implementation, this control variable can take values in a discrete or continuous set. To be consistent with the LTI assumption, only one value can be utilized during the considered signal transmission and it is selected before the transmission is initiated. Since the element is much smaller than the wavelength, it can be modeled as a passive electric circuit. The passiveness implies there is no added noise within the circuit \cite{Renzo2020b}.
In principle, one can also build an RIS with active circuit components (e.g., to make the operation dependent on the content of the impinging signal) but it will inevitably add noise and will not be covered in this tutorial.

The upper part of Fig.~\ref{fig:basic_example}(b) shows the frequency response for the element implementation considered in \cite{Abeywickrama2020a}. The intended carrier frequency is $f_c=3$ GHz and since the frequency response (i.e., reflection coefficient) is complex, we show the phase and amplitude responses around the carrier. Each configuration results in one curve and is achieved by tuning the impedance. In this example, it is tuned by varying the capacitance using a varactor but other implementations use PIN diodes, microelectromechanical systems (MEMS), or optical mechanisms \cite{Tsilipakos2020a,He2019a}.
The phase response is shown for four different capacitance values, which have been selected to give the phase-shifts $\pi/2,0,-\pi/2,\pi$ at the carrier frequency. There are large phase variations over the GHz range, created by the linear phase-shift that a constant time delay would produce and non-linearity created by the frequency-dependent impedance of the element.
However, we can neglect the latter effect if the signal bandwidth $B$ is limited to a few tens of MHz.
The frequency responses of wireless channels are typically varying faster or equally fast with the frequency, thus it is usually these channels that determine whether the end-to-end channel $h(t)$ is narrowband or wideband.
The phase-shifts are caused by three phenomena. The example curves  begin close to $+\pi$ because the reradiated electric field is inverted. As $f$ increases, a constant time delay leads to a larger phase-shift.

The amplitude response is also shown in Fig.~\ref{fig:basic_example}(b) and reveals that the amplitude loss depends on both the frequency and capacitance. The losses are largest when tuning the RIS to achieve zero phase response due to resonance in the circuit.
However, a few dB of signal losses in the RIS element is a minor issue compared to the propagation losses over wireless channels that can be at the order of 100 dB.

The signal that is reradiated from element $n$ propagates to the receiver over an LTI channel with an arbitrary impulse response $b_{n,\pb}(t)$. Since the transmitted signal propagates via element $n$ over a cascade of three LTI filters, the joint impulse response is the convolution of their impulse responses:
$(b_{n,\pb} * \vartheta_{n,\pb; \theta_n} * a_{n,\pb})(t)$.
This happens for all the $N$ elements, thus we obtain the input-output relation
\begin{equation} \label{eq:input-output-passband-RIS}
y_{\pb}(t) = \sum_{n=1}^{N} (b_{n,\pb} * \vartheta_{n,\pb; \theta_n} * a_{n,\pb} *x_{\pb})(t)  = \Bigg( \underbrace{\left[ \sum_{n=1}^{N} b_{n,\pb} * \vartheta_{n,\pb; \theta_n} * a_{n,\pb} \right]}_{{}= h_{\pb; \vect{\theta}}} * x_{\pb} \Bigg)(t)
\end{equation}
where we identify  $h_{\pb;\vect{\theta}}(t) =  \sum_{n=1}^{N} (b_{n,\pb} * \vartheta_{n,\pb; \theta_n} * a_{n,\pb} ) (t)$
as the impulse response of the end-to-end system.
Recall from~\eqref{eq:input-output-baseband} that filtering in the passband can be transformed into pseudo-baseband filtering  by downshifting the filters. By applying this principle to each filter in $h_{\pb;\vect{\theta}}(t)$, we obtain the complex pseudo-baseband representation \vspace{-6mm}
\begin{equation} \label{eq:input-output-baseband-RIS}
y(t) = \sum_{n=1}^{N} (b_{n} * \vartheta_{n; \theta_n} * a_{n} *x)(t) 
\end{equation}
where $a_n(t) = a_{n,\pb}(t) e^{-\imagunit 2\pi f_c t}$, $b_n(t) = b_{n,\pb}(t) e^{-\imagunit 2\pi f_c t}$, and $\vartheta_{n; \theta_n}(t) = \vartheta_{n,\pb; \theta_n}(t) e^{-\imagunit 2\pi f_c t}$ are the channels and filter associated with element $n$.
The end-to-end channel has impulse response
$h_{\vect{\theta}}(t)= \sum_{n=1}^{N} (b_{n} * \vartheta_{n; \theta_n} * a_{n})(t)$.
The fact that the convolution between a chain of impulse responses in the passband becomes a convolution between the corresponding chain of pseudo-baseband impulse responses is a unique feature of the complex pseudo-baseband representation that we consider.
The conventional textbook formulation where each filter is assumed to be passband gives rise to extra scaling factors.

\vspace{-4mm}

\subsection{Equivalent discrete-time system model}

The continuous-time complex baseband signal $x(t)$ is usually generated to represent a complex discrete-time signal $x[m]$, where $m$ is the integer index, via pulse-amplitude modulation (PAM). 
We consider ideal PAM using a unit-energy sinc-pulse $p(t) = \sqrt{B}\sinc(Bt)$ and the symbol rate $B$, for which
\begin{equation} \label{eq:PAM}
x(t) = \sum_{m=-\infty}^{\infty} x[m] p \left( t- \frac{m}{B} \right).
\end{equation}
Since the actual input signal $x[m]$ is in discrete time, it is convenient to abstract away the entire continuous-time description by sampling the received signal to obtain an end-to-end discrete-time system model. Before sampling, we must add the thermal receiver noise and lowpass filtering at the receiver into the model. We model the noise by a white circularly symmetric complex Gaussian random process  $w(t)$ with power-spectral density $N_0$.
Adding it to the received signal in \eqref{eq:input-output-baseband} as $z(t) = y(t) + w(t)$, we obtain
\begin{equation}
z(t) = (h_{\vect{\theta}} *x)(t) + w(t) = \sum_{m=-\infty}^{\infty} x[m] \, (h_{\vect{\theta}} *p ) \left( t- \frac{m}{B}  \right) + w(t)
\end{equation}
where the equality follows from \eqref{eq:PAM}. Since the desired signal is bandlimited to $|f| \leq B/2$, while the noise is not,  we filter $z(t)$  using an ideal lowpass filter with impulse response $p(t)$, the same as in \eqref{eq:PAM}, to remove the out-of-band noise. 
We then take samples at the symbol rate, at time instants $t=k/B+\eta$ where $k$ is the integer sample index and $\eta$ is the sampling delay at the receiver, to obtain
\begin{equation} \label{eq:symbol-sampled-model}
z[k] = (p*z)(t) \Big|_{t=k/B+\eta} =   \sum_{m=-\infty}^{\infty} x[m] h_{\vect{\theta}}[k-m] + w[k]
\end{equation}
where the discrete-time impulse response is defined as
\begin{equation} \label{eq:end-to-end-impulse-response}
h_{\vect{\theta}}[k] = (p * h_{\vect{\theta}} * p) \left(  t \right) \Big|_{t=k/B+\eta} = \sum_{n=1}^{N} (p * b_{n} * \vartheta_{n; \theta_n} * a_{n} * p) (t) \Big|_{t=k/B+\eta}  
\end{equation}
by inserting the RIS system model from \eqref{eq:input-output-baseband-RIS}. Note that the discrete-time impulse response is created by lowpass filtering of the end-to-end continuous-time impulse response $h_{\vect{\theta}}(t)$ and then taking samples of it. The discrete-time noise $w[k]$ in \eqref{eq:symbol-sampled-model} is circularly symmetric complex Gaussian distributed (since $w(t)$ is Gaussian) and independent for different $k$ (since $(p*w)(t)$ has $\sinc(B (t_1-t_2))$ as autocorrelation):
\begin{equation}
w[k] = (p*w) \left( t \right)  \Big|_{t=k/B+\eta}  \sim \CN(0,N_0).
\end{equation}

The discrete-time model in \eqref{eq:symbol-sampled-model} applies to any system but can be simplified by considering the specific properties that wireless channels and practical signals and systems possess: 1) The channels are causal and incur a finite maximum delay; 2) The ideal sinc pulse $p(t)$ is approximated by a time-limited Nyquist pulse (i.e., requiring a bandwidth slightly larger than $B$, where $B$ is the symbol rate); and 3) the sampling delay $\eta$ is selected to obtain a causal discrete-time system. This implies that the channel is a finite impulse response (FIR) filter with $M \geq 1$ terms:
\begin{equation} \label{eq:symbol-sampled-model-FIR}
z[k] =  \sum_{m=k-M+1}^{k} x[m] h_{\vect{\theta}}[k-m] + w[k] = \sum_{\ell=0}^{M-1} h_{\vect{\theta}}[\ell] x[k-\ell] + w[k]
\end{equation}
where $h_{\vect{\theta}}[0],\ldots,h_{\vect{\theta}}[M-1]$ are the non-zero components of the impulse response. \vspace{-5mm}

\subsection{Canonical multicarrier system model}

The discrete-time system model in \eqref{eq:symbol-sampled-model-FIR} describes a dispersive channel with a memory of $M-1$ previous symbols; that is, the received $z[k]$ contains not only the currently transmitted signal $x[k]$ but also intersymbol interference from $x[k-1],\ldots,x[k-M+1]$. A common way to untangle the interference is to 
design the transmitted symbols using orthogonal frequency-division multiplexing (OFDM), transforming the channel into a collection of separate frequency subcarriers. We will provide the corresponding reformulated system model, which will be utilized for both communication and localization.

Suppose we want to transmit a block of $K$ symbols, $\chi[0],\ldots,\chi[K-1]$, and append a so-called cyclic prefix to obtain the following sequence of length $K+M-1$  that can be transmitted over the input-output system defined in \eqref{eq:symbol-sampled-model-FIR}:
\begin{equation}
x[k] = \begin{cases}
\chi[k] & k = 0,\ldots,K-1 \\
\chi[k+K] & k=-M+1,\ldots,-1.
\end{cases}
\end{equation}
Since we added the last $M-1$ symbols as a prefix, 
we can interpret \eqref{eq:symbol-sampled-model-FIR} as a cyclic convolution between $\{ \chi[k] : k=0,\ldots,K-1\}$ and $\{h_{\vect{\theta}}[k] : k = 0,\ldots,M-1\}$, plus noise, if $K>M$. Let us define the $K$-point discrete Fourier transform (DFT) of an arbitrary sequence $s[k]$ as $\mathcal{F}_d\{ s[k] \} = \frac{1}{\sqrt{K}} \sum_{k=0}^{K-1} s[k] e^{-\imagunit 2 \pi k \nu /K}$, where the scaling factor keeps the energy constant.
Taking the discrete Fourier transform of \eqref{eq:symbol-sampled-model-FIR} and utilizing that cyclic convolution becomes the product of the corresponding Fourier transforms, we obtain the $K$ orthogonal subcarriers
\begin{equation} \label{eq:symbol-sampled-model-OFDM}
\bar{z}[\nu] = \bar{h}_{\vect{\theta}}[\nu] \bar{x}[\nu] + \bar{w}[\nu], \quad \nu = 0, \ldots, K-1
\end{equation}
where $\bar{z}[\nu] = \mathcal{F}_d \{ z[k] \}$ and $\bar{x}[\nu] = \mathcal{F}_d \{ x[k] \}$ describe the received and transmitted signals, respectively, in the frequency domain. 
At subcarrier $\nu$, the frequency response of the end-to-end channel is
\begin{equation} \label{eq:frequency-response}
\bar{h}_{\vect{\theta}}[\nu] = \sum_{k=0}^{M-1} h_{\vect{\theta}}[k] e^{-\imagunit 2 \pi k \nu /K}
\end{equation}
and the transformed noise $\bar{w}[\nu] = \mathcal{F}_d \{ w[k] \} \sim \CN(0,N_0)$ is independent for $\nu=0,\ldots,K-1$. 
Notice that \eqref{eq:symbol-sampled-model-OFDM} has a more convenient structure than \eqref{eq:symbol-sampled-model-FIR} since there is no intersymbol interference. It is known as a discrete memoryless channel with additive white Gaussian noise (AWGN). OFDM exploits this feature by treating $\bar{x}[\nu]$ as the transmitted signal and $\bar{z}[\nu]$ as the received signal. In an OFDM implementation, the transmitted time-domain $x[k]$ is generated from $\bar{x}[\nu]$ by an inverse Fourier transform, while the receiver computes the Fourier transform of its received signal $z[k]$ to obtain $\bar{z}[\nu]$.

\subsection{Example of multipath channels}

We will now give a concrete example of how the end-to-end channel $\bar{h}_{\vect{\theta}}[\nu]$ in the OFDM system model \eqref{eq:symbol-sampled-model-OFDM} is determined by the propagation channels and RIS elements. 
Suppose the channel from the transmitter to the $n$th RIS element consists of $L_a$ propagation paths, then the impulse response is modeled as
\begin{equation}
a_{n,\pb}(t) = \sum_{l=1}^{L_a} \sqrt{\alpha_n^{l}} \delta(t- \tau_{n,a}^{l}) \quad \Rightarrow \quad a_{n}(t) = \sum_{l=1}^{L_a} \sqrt{\alpha_n^{l}} e^{-\imagunit 2\pi f_c t} \delta(t- \tau_{n,a}^{l})
\end{equation}
where $\alpha_n^{l} \in [0,1]$ is the propagation loss and $ \tau_{n,a}^{l}\geq 0$ is the delay of the $l$th path, while $\delta(t)$ denotes the Dirac delta function. Similarly, suppose there are $L_b$ propagation paths from the $n$th RIS element to the receiver, then the impulse response can be modeled as
\begin{equation}
b_{n,\pb}(t) = \sum_{\ell=1}^{L_b} \sqrt{\beta_n^{\ell}} \delta(t- \tau_{n,b}^{\ell}) \quad \Rightarrow \quad b_{n}(t) = \sum_{\ell=1}^{L_b}\sqrt{\beta_n^{\ell}} e^{-\imagunit 2\pi f_c t} \delta(t- \tau_{n,b}^{\ell})
\end{equation}
where $\beta_n^{\ell} \in [0,1]$ is the propagation loss and $ \tau_{n,b}^{\ell}\geq 0$ is the delay of the $\ell$th path. 
We assume the signal bandwidth is sufficiently small to 
make the frequency response of the RIS element constant in amplitude and time delay (i.e., linear phase); that is, the RIS is narrowband while the wireless channels might be wideband. For a given configuration $\theta_n \in \Omega$, selected from some set $\Omega$ of feasible configurations, element $n$ is reradiating a fraction $\gamma_{\theta_n}  \in [0,1]$ of the incident signal power and incurs a delay of $\tau_{\theta_n} \geq 0$, so that
\begin{equation} \label{eq:ris-element-impulse-response}
\vartheta_{n,\pb; \theta_n}(t) = \sqrt{\gamma_{\theta_n}} \delta(t- \tau_{\theta_n})  \quad \Rightarrow \quad
 \vartheta_{n; \theta_n}(t) =  \sqrt{\gamma_{\theta_n}} e^{-\imagunit 2\pi f_c t} \delta(t- \tau_{\theta_n}).
\end{equation}
Under these assumptions, the discrete-time impulse response in \eqref{eq:end-to-end-impulse-response} particularizes to
\begin{align} 
h_{\vect{\theta}}[k] &= \sum_{n=1}^{N}  \sum_{l=1}^{L_a}  \sum_{\ell=1}^{L_b} \sqrt{\alpha_n^{l} \beta_n^{\ell} \gamma_{\theta_n}} e^{-\imagunit 2\pi f_c (\tau_{n,a}^{l}+\tau_{n,b}^{\ell}+\tau_{\theta_n})}
\underbrace{\sinc \big( k + B( \eta -\tau_{n,a}^{l}-\tau_{n,b}^{\ell}-\tau_{\theta_n}) \big)}_{\approx \sinc \big( k + B( \eta -\tau_{n,a}^{l}-\tau_{n,b}^{\ell}) \big)} \label{eq:example-end-to-end}
\end{align}
where the approximation utilizes the fact that the delay in the RIS is much smaller than the propagation delays, so its impact on the symbol rate is negligible: $B \tau_{\theta_n} \approx 0$.
However, since $f_c \gg B$, the RIS creates phase-shifts $2\pi f_c \tau_{\theta_n}$ in \eqref{eq:example-end-to-end} that are substantial (within a few periods of $2\pi$), as illustrated in Fig.~\ref{fig:basic_example}(b).

We notice from \eqref{eq:example-end-to-end} that there are $N L_a L_b$ paths from the transmitter to the receiver, each having a unique propagation loss $\alpha_n^{l} \beta_n^{\ell} \gamma_{\theta_n}$ which is the product of the losses between the transmitter to an RIS element, inside the element, and from the element to the receiver. Due to the product operation, each path is very weak but the large number of paths can potentially lead to a good SNR.
Each path is also associated with a phase-shift $e^{-\imagunit 2\pi f_c (\tau_{n,a}^{l}+\tau_{n,b}^{\ell}+\tau_{\theta_n})}$ containing the accumulated delays.
The sinc-function determines how the signal energy carried by the path is divided between the $M$ taps of the FIR filter.

The frequency response $\bar{h}_{\vect{\theta}}[\nu]$ can now be computed using \eqref{eq:frequency-response}. To obtain a compact expression, we first notice that \eqref{eq:example-end-to-end} can be expressed as an inner product of two vectors:
\begin{align} 
h_{\vect{\theta}}[k] &= 
\underbrace{\begin{bmatrix}
\sum\limits_{l=1}^{L_a}  \sum\limits_{\ell=1}^{L_b} \sqrt{\alpha_1^{l} \beta_1^{\ell} } e^{-\imagunit 2\pi f_c (\tau_{1,a}^{l}+\tau_{1,b}^{\ell})}
\sinc \big( k + B( \eta -\tau_{1,a}^{l}-\tau_{1,b}^{\ell}) \big) \\
\vdots \\
\sum\limits_{l=1}^{L_a}  \sum\limits_{\ell=1}^{L_b} \sqrt{\alpha_N^{l} \beta_N^{\ell} } e^{-\imagunit 2\pi f_c (\tau_{N,a}^{l}+\tau_{N,b}^{\ell})}
\sinc \big( k + B( \eta -\tau_{N,a}^{l}-\tau_{N,b}^{\ell}) \big)
\end{bmatrix}^{\Ttran}}_{\vect{v}_k^{\Ttran}}
\underbrace{\begin{bmatrix}
\sqrt{\gamma_{\theta_1}} e^{-\imagunit 2\pi f_c \tau_{\theta_1}} \\
\vdots \\
\sqrt{\gamma_{\theta_N}} e^{-\imagunit 2\pi f_c \tau_{\theta_N}}
\end{bmatrix}}_{\vect{\omega}_{\vect{\theta}}}.
\label{eq:example-end-to-end-vectors}
\end{align}
The propagation channels determine $\vect{v}_k \in \mathbb{C}^N$, while the RIS determines $\vect{\omega}_{\vect{\theta}}  \in \mathbb{C}^N$ and it is the same for all $k$.
Hence, $\vect{v}_k$ is given by nature while $\vect{\omega}_{\vect{\theta}}$ is controllable.
We can compute the frequency response as
\begin{equation} \label{eq:OFDM-response}
\underbrace{\begin{bmatrix}
\bar{h}_{\vect{\theta}}[0] \\
\vdots \\
\bar{h}_{\vect{\theta}}[K-1]
\end{bmatrix}}_{\bar{\vect{h}}_{\vect{\theta}}}
= \vect{F} \underbrace{\begin{bmatrix}
h_{\vect{\theta}}[0] \\
\vdots \\
h_{\vect{\theta}}[M-1]
\end{bmatrix}}_{\vect{h}}
= \vect{F} \vect{V}^{\Ttran} \vect{\omega}_{\vect{\theta}}
\end{equation}
where $\vect{V}= [\vect{v}_0, \ldots, \vect{v}_{M-1}]$ is an $N \times M$ matrix and $\vect{F}$ is a $K \times M$ DFT matrix with the $(\nu,k)$th element being $e^{-\imagunit 2 \pi k \nu / K}$. 
We will make use of this notation when considering wideband systems. \vspace{-2mm}

\subsection{Simplified narrowband system model}

When there is only one strong path to and from the RIS (i.e., the LOS paths), we can select the sampling delay $\eta$ to make $B(\eta -\tau_{n,a}^{l}-\tau_{n,b}^{\ell}) = 0$ for that path.
By setting $L_a=L_b = 1$ and omitting the superscripts indicating the path indices, we can then rewrite the impulse response in \eqref{eq:example-end-to-end} as
\begin{equation}
h_{\vect{\theta}}[k] = \sum_{n=1}^{N} \sqrt{\alpha_n \beta_n \gamma_{\theta_n}} e^{-\imagunit 2\pi f_c (\tau_{n,a}+\tau_{n,b}+\tau_{\theta_n})}
\sinc \left( k \right) = \begin{cases}  \sum_{n=1}^{N} \sqrt{\alpha_n \beta_n \gamma_{\theta_n}} e^{-\imagunit 2\pi f_c (\tau_{n,a}+\tau_{n,b}+\tau_{\theta_n})} & k=0 \\
0 & k \neq 0.
\end{cases}
\end{equation}
A channel of this kind is called narrowband and the input-output system in \eqref{eq:symbol-sampled-model} simplifies to
\begin{equation} \label{eq:symbol-sampled-model-narrowband}
z[k] = h_{\vect{\theta}}[0] x[k] + w[k] = \sum_{n=1}^{N} \sqrt{\alpha_n \beta_n \gamma_{\theta_n}} e^{-\imagunit 2\pi f_c (\tau_{n,a}+\tau_{n,b}+\tau_{\theta_n})}   x[k]  + w[k].
\end{equation}
This is a popular special case where there is no need for OFDM since there is no intersymbol interference.
The model can also be derived when having multiple paths 
with a delay spread much smaller than the sampling period $1/B$, making them indistinguishable.
In the LOS case, we can relate the channel coefficients to the steering vector $\vect{a}(\vect{\phi})\in \mathbb{C}^N$ of the RIS, which describes the phase-shift pattern over the elements when a plane wave arrives from the azimuth/elevation angle pair $\vect{\phi} \in \mathbb{R}^2$.
The RIS geometry determines the steering vector and a general way to compute it is found in \cite{Bjornson2021a}.
Using the steering vector, we can write
$\sum_{n=1}^{N} \sqrt{\alpha_n \beta_n \gamma_{\theta_n}} e^{-\imagunit 2\pi f_c (\tau_{n,a}+\tau_{n,b}+\tau_{\theta_n})} = 
\sqrt{\alpha \beta} e^{j \psi_{\text{RIS}}}
(\vect{a}(\vect{\phi}_a) \odot \vect{a}(\vect{\phi}_b))^{\Ttran}
\vect{\omega}_{\vect{\theta}}$, where $\alpha= \alpha_n, \beta=\beta_n$ for all $n$, $\psi_{\text{RIS}}$ is a common phase shift, $\vect{\phi}_a$ is the angle to the transmitter, and $\vect{\phi}_b$ is the angle to the receiver. 
We will use this geometric modeling for simulations and localization.

\section{RIS-aided communications}
\label{sec:RIScomms}

The central element when characterizing the performance of a communication channel is the probabilistic relation between the discrete-time input $x$ and output $z$, specified by the conditional probability density function (PDF) $p(z|x)$. For example, the narrowband system in~\eqref{eq:symbol-sampled-model-narrowband} has the complex symbols $x$ and $z$ as inputs and outputs, respectively. The received signal for the $k$th transmitted symbol is 
\begin{equation} \label{eq:GaussianBaseband}
z[k] = h_{\vect{\theta}} \, x[k] + w[k], \quad k=0,1,\ldots
\end{equation}
where a discrete sequence $\{x[k]\}$ of codeword symbols describing the payload data is transmitted and each one is attenuated by a factor $h_{\vect{\theta}} \in \mathbb{C}$ and corrupted by the independent noise $w[k]\sim \CN(0,N_0)$. To arrive at the well-known communication model of an AWGN channel, we need some further assumptions. First, $h_{\vect{\theta}}$ is fixed for all symbols in a given codeword, and its value is known to the transmitter and receiver. Second, we assume that the power limit of the sender is  $P$ Watt. Since there are $B$ symbols per second, each symbol should satisfy the power constraint $\mathbb{E} \{ | x[k] |^2\} \leq P/B$. Finally, the transmitter must know the SNR of the channel, given by $\mathsf{SNR}  = P | h_{\vect{\theta}} |^2 / (B N_0)$. The capacity of this AWGN channel is
\begin{equation} \label{eq:capacity-AWGN}
    C = B \log_2 \left( 1+ \mathsf{SNR} \right) = B \log_2 \left( 1+ \frac{P | h_{\vect{\theta}} |^2}{B N_0} \right) \quad \textrm{bit/s}.
\end{equation}
If any of the above assumptions are violated (e.g., $h_{\vect{\theta}}$ is not known by the receiver or the SNR is not known by the transmitter), the channel is not AWGN and the capacity formula \eqref{eq:capacity-AWGN} is not valid. For a deeper discussion on how assumptions affect the definition of a communication channel, see~\cite[Ch.~6]{Popovski2020a}. 

This discussion can be extrapolated to the OFDM channel in \eqref{eq:symbol-sampled-model-OFDM}, obtained as a superposition of $K$ parallel memoryless AWGN channels. In this case, we need to consider a set of $K$ channel values $\{\bar{h}_{\vect{\theta}}[\nu]\}$, each of them associated with one of the narrowband subcarrier channels. 

Given the transmitter, the receiver, and a narrowband channel, an obvious objective of an RIS would be to select the configuration ${\vect{\theta}}$
to create a channel $h_{\vect{\theta}}$ that maximizes the capacity \eqref{eq:capacity-AWGN}. More generally, in case of OFDM, the objective is to create a set of $K$ channels $\{\bar{h}_{\vect{\theta}}[\nu]\}$ that maximizes the sum of the capacity of the constituent subcarrier channels. In this case, the values of the channels $\bar{h}_{\vect{\theta}}[0],\ldots,\bar{h}_{\vect{\theta}}[K-1]$ may not be independently optimized, as they are determined by the same RIS configuration.

 To have the RIS configured, it is necessary that the RIS is capable to receive control information from the radio infrastructure.
There are two principal types of channels for sending this control information: out-of-band and in-band. An \emph{out-of-band control channel}
does not consume part of the useful bandwidth $B$ and is implemented as a wired link or a wireless channel that uses different frequency spectrum. The RIS-controlled channel in \eqref{eq:GaussianBaseband} implicitly assumes that the control information has been exchanged through an out-of-band channel before the actual communication starts.

In contrast, an \emph{in-band control channel} consumes part of the useful bandwidth to configure the RIS and this should be factored in when computing the overall capacity of the wireless channel. The in-band control information can be sent before the actual communication. However, it is also possible to have an in-band control channel in which the RIS control information is sent simultaneously with the payload data. In this case, the RIS should be able to decode the control information embedded into $x[k]$ and based on that, causally change the value of $h_{\vect{\theta}}$ (i.e., change $\vect{\theta}$) for symbols with indices $j>k$. It is immediately clear that the end-to-end channel cannot be an AWGN channel anymore. In an information-theoretic sense, this situation corresponds to a relay channel in which the source broadcasts two types of data: payload data (intended for the receiver) and control data (intended for the RIS that becomes a relay node). Based on the received data, the RIS changes the configuration of the end-to-end channel.

\begin{figure}[t!]
	\centering 
	\begin{overpic}[width=\textwidth,tics=10]{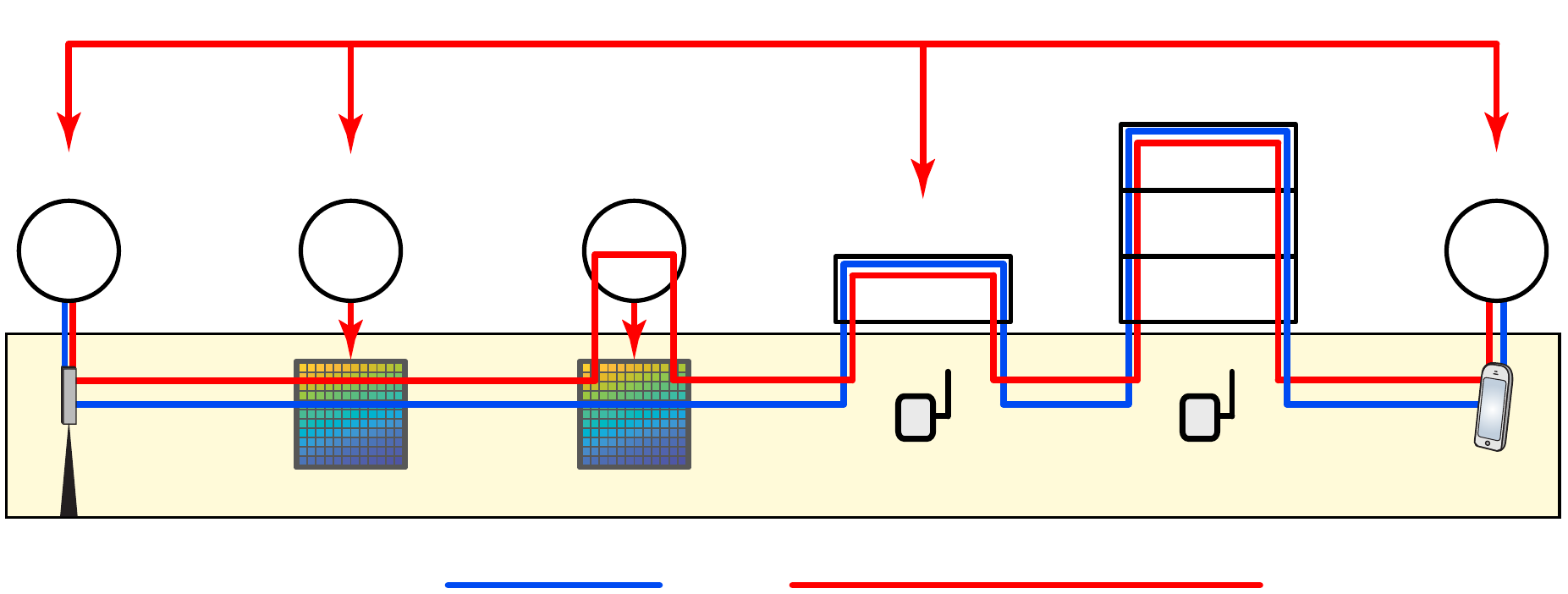}
		\put(29.5,1.5){Payload data}
		\put(51,1.5){Control information or action}
		\put(47,6){Layer 0}
  \put (55.5,18) {Layer 1}
  \put (73.5,18.5) {Layer 1}
  \put (73.5,23) {Layer 2}
  \put (73.5,26.5) {Layer 3}
  \put (1,26) {Node 1}
  \put (20,26) {RIS 1}
  \put (38,26) {RIS 2}
  \put (54.5,23) {AF relay}
  \put (73,31) {DF relay}
  \put (92,26) {Node 2}
  \put (36,36) {Out-of-band control channel}
\end{overpic}  \vspace{-4mm}
	\caption{A layered perspective on RIS and relay functionality. The wireless connection between Node 1 and Node 2 is supported by a cascade of RIS 1 with out-of-band control, RIS 2 with in-band control, AF relay and DF relay.}  \vspace{-3mm}
	\label{fig:LayeredModel}  
\end{figure}

The works~\cite{DiRenzoOJ20,Bjornson2020a,BS20} compare RIS-aided and traditional relay-aided systems. Here we provide a different perspective, using the layering framework, as depicted in Fig.~\ref{fig:LayeredModel}. 
We consider a wireless connection between Node 1 and Node 2, aided by a cascade of RISs and relays. Nodes 1 and 2 implement all protocol layers (not depicted). Suppose Node 1 transmits to Node 2. The wireless signal is reflected by RIS 1, which is configured using an out-of-band control channel. The payload data (blue line) stays at Layer 0 (the wireless medium), meaning that the RIS only affects the propagation environment without processing the communication flow. The control information (red line) shows that there needs to be an out-of-band communication between Node 1, Node 2, and RIS 1 to select a suitable configuration $\vect{\theta}$. Next, the signal reaches RIS 2, which uses an in-band control channel. The payload data stays at Layer 0, while the RIS controller decodes the control information and adapts its configuration.
Next, the signal reaches a non-regenerative amplify-and-forward (AF) relay. The payload data goes through Layer 1 (physical) where the signal is amplified and the active circuitry introduces additional noise. The depicted control channel is out-of-band but, similar to the RIS 1 case, it can be implemented in-band, using a dedicated communication protocol that is not used by the payload data. Finally, the decode-and-forward (DF) relay decodes both payload and control data, and is capable to interpret the control information.

\subsection{RIS design for narrowband capacity maximization}\label{SectionIIIA}

To explain how an RIS can be used to maximize the capacity, 
we begin by considering a simple setup: a single-antenna transmitter communicates with a single-antenna receiver over a narrowband channel. We further assume that the RIS elements can be perfectly configured: we can select $\tau_{\theta_n} \geq 0$ so that $2\pi f_c \tau_{\theta_n}$ can take any value between $0$ and $2\pi$ for $n=1,\ldots,N$, while the amplitude response is constant $\gamma_{\theta_n} = \gamma$.

The considered system is a memoryless AWGN channel with $h_{\vect{\theta}} = \sum_{n=1}^{N} \sqrt{\alpha_n \beta_n \gamma} e^{-\imagunit 2\pi f_c (\tau_{n,a}+\tau_{n,b}+\tau_{\theta_n})}$, thus the capacity in \eqref{eq:capacity-AWGN} becomes $B \log_2(1+\mathsf{SNR})$ with
\begin{align} 
    \mathsf{SNR} &= \frac{P}{B N_0} \left| \sum_{n=1}^{N} \sqrt{\alpha_n \beta_n \gamma} e^{-\imagunit 2\pi f_c (\tau_{n,a}+\tau_{n,b}+\tau_{\theta_n})} \right|^2  \label{eq:SNR-vector-description} =  \frac{P}{B N_0} \left|
        \begin{bmatrix}
    \sqrt[4]{\alpha_1 \beta_1 \gamma} \\
    \vdots \\
    \sqrt[4]{\alpha_N \beta_N \gamma} \
    \end{bmatrix}^{\Ttran}
    \begin{bmatrix}
    e^{-\imagunit 2\pi f_c (\tau_{1,a}+\tau_{1,b}+\tau_{\theta_1})} \sqrt[4]{\alpha_1 \beta_1 \gamma} \\
    \vdots \\
    e^{-\imagunit 2\pi f_c (\tau_{N,a}+\tau_{N,b}+\tau_{\theta_N})} \sqrt[4]{\alpha_N \beta_N \gamma} \
    \end{bmatrix}  \right|^2 \\ \label{eq:SNR-vector-description-CS}
& \leq \frac{P}{B N_0} \left| \sum_{n=1}^{N} \sqrt{\alpha_n \beta_n \gamma}  \right|^2
\end{align}
where the last step follows from the Cauchy-Schwartz inequality. The upper bound in that inequality is achieved if and only if the two vectors in \eqref{eq:SNR-vector-description} are parallel, which occurs when $e^{-\imagunit 2\pi f_c (\tau_{n,a}+\tau_{n,b}+\tau_{\theta_n})}$ is the same for all $n$ \cite{Wu2018a}. Hence, each RIS element should phase-shift its reradiated signal so that it reaches the receiver synchronously in phase with the signals from all the other RIS elements. There are multiple solutions due to the phase periodicity, but the solution causing the minimum overall time delay is \cite{Matthiesen2020a}
\begin{equation} \label{eq:optimized-delays-basic}
    \tau_{\theta_n} = \left(\max_{i=1,\ldots,N} \tau_{i,a}+\tau_{i,b} \right) - \tau_{n,a}-\tau_{n,b}
\end{equation}
where $\tau_{\theta_n}=0$ for the element experiencing the largest propagation delay, while all other elements add positive delays $\tau_{\theta_n}>0$ to match the largest propagation delay.

Suppose all $N$ paths have the same propagation loss: $\alpha_n \beta_n = \alpha \beta$. This is a common property when the RIS is in the far-field of the transmitter and receiver. It then follows that $| \sum_{n=1}^{N} \sqrt{\alpha_n \beta_n \gamma}  |^2 = N^2\alpha \beta \gamma$, thus the SNR grows quadratically with the number of RIS elements \cite{Wu2018a,Basar2019b}. The intuition behind this result is that the surface intercepts signal energy proportional to $N$ (i.e., proportional to its area) and then focuses the reradiated signals to increase the received signal energy proportional to $N$ (thanks to constructive interference of the signals from the $N$ elements).
This result indicates that a physically large RIS is much more effective than a small RIS, which is fundamentally important since $N^2$ is  multiplied with $\alpha \beta$, which is the product of two propagation losses that both can be very small numbers.

\begin{figure}[t!]
	\centering  
\vspace{-7mm}
	\begin{overpic}[width=.7\columnwidth,tics=10]{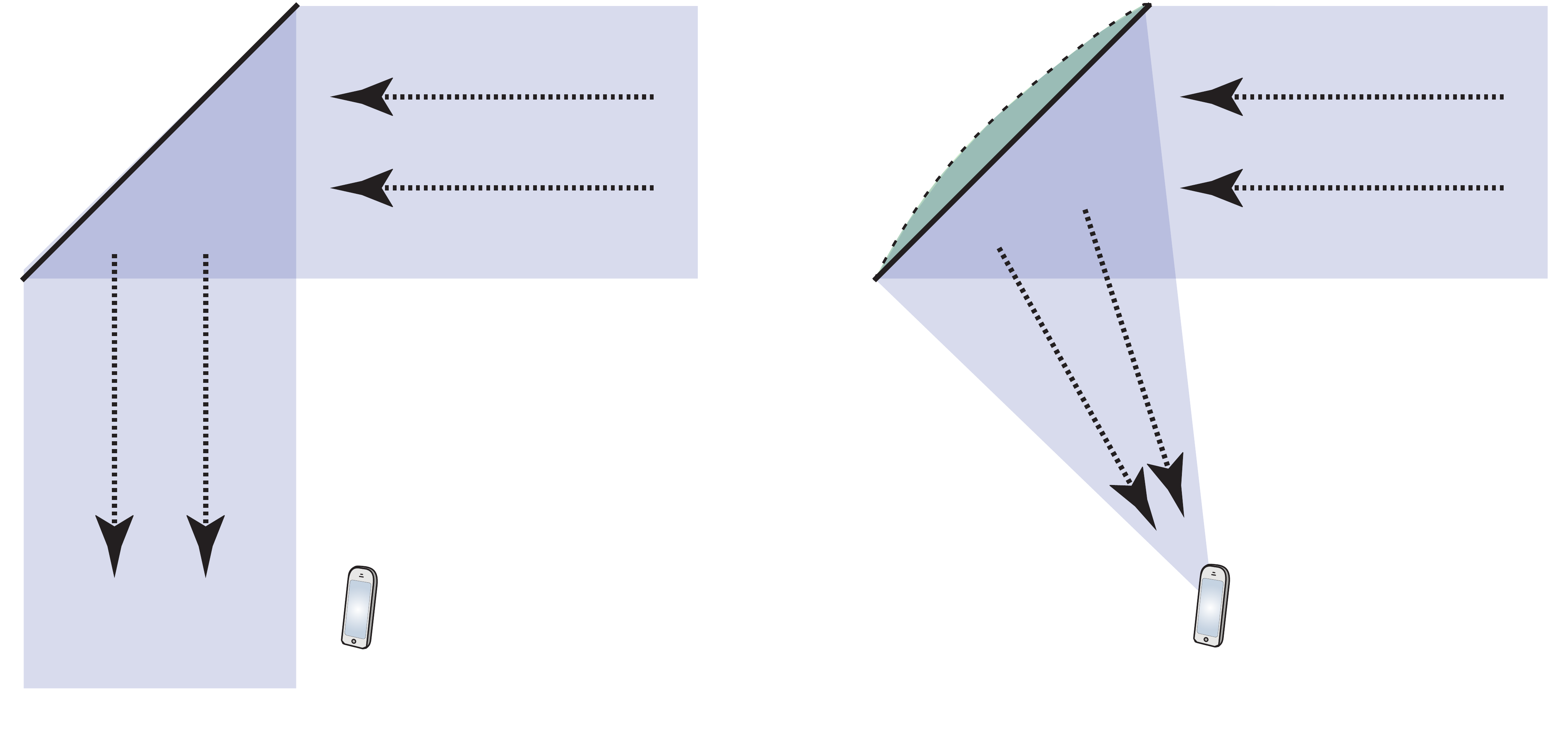}
		\put(-8,44){Conventional}
		\put(-8,40){homogeneous}
		\put(-8,36){surface}
  \put (56,40) {RIS}
    \put (25,5) {Receiver}
      \put (80,5) {Receiver}
\end{overpic}  \vspace{-6mm}
	\caption{A large homogeneous flat surface will reflect an incident plane wave in another direction, determined by Snell's law. Parallel incident rays remain parallel after reflection. In contrast, an RIS of the same physical dimensions can be configured to synthesize the shape of a different object (here: a parabolic reflector), thereby controlling the direction and shape of the reflected waveform.} \vspace{-5mm}
	\label{figure:geometrical_example}  
\end{figure}

A geometrical interpretation of the optimal RIS configuration is provided in Fig.~\ref{figure:geometrical_example}, where a plane wave is incident on a large flat surface. If it is a homogeneous metal surface, as shown to the left, the plane wave changes direction according to Snell's law but otherwise is unaffected (e.g., two rays remain parallel). Each point on the surface reradiates the incident signal without causing any extra delays.
In the illustrated scenario, the reflected signal does not reach the receiver. 
If the surface is replaced by an RIS, as shown to the right, the optimized configuration focuses the reradiated signal at the receiver. The configuration in \eqref{eq:optimized-delays-basic} {adds extra delays in the center of the surface to make the propagation time to the receiver equal for all parallel rays that are reflected. As illustrated by the dashed line, the RIS is synthesizing how signals would have been reflected by a parabolic surface, where the length of each path via the surface is equal.}
This configuration will both change the shape and the main direction of the waveform (e.g., two parallel incident rays have different directions when reradiated). The same effect could have been mechanically achieved by rotating and bending the flat metal surface, but doing it electronically using an RIS adds great flexibility since different surface shapes can be synthesized at different times.

\subsection{Narrowband capacity maximization with a partially uncontrollable channel}

In practice, there are likely propagation paths between the transmitter and the receiver not involving the RIS, thus outside its control. Recall from Fig.~\ref{fig:basic_example}(b) that these paths constitute the uncontrollable channel with impulse response $h_{d,\pb}(t)$.
In the narrowband case, it can be represented in the discrete-time complex pseudo-baseband by an impulse response $\sqrt{\rho} e^{-\imagunit 2\pi f_c \tau_d}$, where $\rho  \in [0,1]$ is the propagation loss and $\tau_d \geq 0$ is the delay.
We then obtain a memoryless AWGN channel with $h_{\vect{\theta}}  = \sqrt{\rho} e^{-\imagunit 2\pi f_c \tau_d}+ \sum_{n=1}^{N} \sqrt{\alpha_n \beta_n \gamma} e^{-\imagunit 2\pi f_c (\tau_{n,a}+\tau_{n,b}+\tau_{\theta_n})}$,
for which the capacity in \eqref{eq:capacity-AWGN} becomes $B \log_2(1+\mathsf{SNR})$ with
\begin{align} \nonumber
    \mathsf{SNR} &= \frac{P}{B N_0} \left|  \sqrt{\rho} e^{-\imagunit 2\pi f_c \tau_d} + \sum_{n=1}^{N} \sqrt{\alpha_n \beta_n \gamma} e^{-\imagunit 2\pi f_c (\tau_{n,a}+\tau_{n,b}+\tau_{\theta_n})} \right|^2 \\
    & \leq \frac{P}{B N_0} \left|  \sqrt{\rho} e^{-\imagunit 2\pi f_c \tau_d}+ \sum_{n=1}^{N} \sqrt{\alpha_n \beta_n \gamma} e^{-\imagunit 2\pi f_c \tau_d} \right|^2 = \frac{P}{B N_0} \left|  \sqrt{\rho} + \sum_{n=1}^{N} \sqrt{\alpha_n \beta_n \gamma} \right|^2. \label{eq:SNR-max-direct-channel}
\end{align}
The upper bound is once again obtained by the Cauchy-Schwartz inequality, with the key difference that we cannot control the phase of the uncontrollable channel component. Hence, we need to select the delays of the RIS elements so that the $N$ reradiated signals reach the receiver in phase with the signal over the uncontrollable channel. Note that $e^{-\imagunit 2\pi f_c \tau_d} = e^{-\imagunit 2\pi f_c (\tau_{n,a}+\tau_{n,b}+\tau_{\theta_n})}$ holds if $\tau_{\theta_n} = \tau_d - \tau_{n,a}-\tau_{n,b}$ for $n=1,\ldots,N$, but this results in a negative delay if the uncontrollable channel path is shorter than the paths via the RIS, which is usually the case. Hence, to achieve a causal system implementation, we need to select the delays as
$\tau_{\theta_n} = \left(\tau_d - \tau_{n,a}-\tau_{n,b} \right) + \frac{k_n}{f_c}$
where $k_n$ is an integer such that $\tau_{\theta_n}\geq 0$; that is, $k_n \ge \lceil f_c \left( \tau_{n,a} + \tau_{n,b} - \tau_d \right) \rceil$, where $\lceil \cdot \rceil$ is the ceiling function.
Then, the delay spread $T_\mathrm{d}$ is
\begin{equation}
	T_\mathrm{d} = \max_{i=1,\ldots,N}\{ \tau_{i,a} + \tau_{i,b} + \tau_{\theta_i} \} - \tau_d = \max_i \left\{ \frac{k_i}{f_c} \right\} \ge \max_i \left\{ \frac{1}{f_c} \lceil f_c \left( \tau_{i,a} + \tau_{i,b} - \tau_d \right) \rceil \right\},
\end{equation}
which is minimized for the smallest integer that satisfies the constraint above \cite{Matthiesen2020a}.

The upper bound in \eqref{eq:SNR-max-direct-channel} is achieved when the RIS configuration $\vect{\theta}$ can be selected from a continuous set. Suppose the RIS hardware restricts us to select $\tau_{\theta_n}$ from a discrete set such that $2\pi f_c \tau_{\theta_n} \in \{\pi/2,0,-\pi/2,-\pi\}$, as exemplified in Fig.~\ref{fig:basic_example}(b). The capacity maximization is now a combinatorial problem with $4^N$ possible configurations. Evaluating all options is computationally very complex, but a good heuristic is to rotate each term $e^{-\imagunit 2\pi f_c (\tau_{n,a}+\tau_{n,b}+\tau_{\theta_n})}$ so it is as close to $e^{-\imagunit 2\pi f_c \tau_d}$ as possible. This leads to a partially coherent addition of the $N+1$ components of the channel $h_{\vect{\theta}}$.
One can prove that the SNR loss is only around $8/\pi^2 = -0.9$ dB when having these four configurations \cite{Wu2020b}, which implies that a small number of configurations per element is sufficient when implementing an RIS.

\begin{figure}
	\centering \vspace{-12mm}
	\begin{overpic}[width=.5\columnwidth,tics=10]{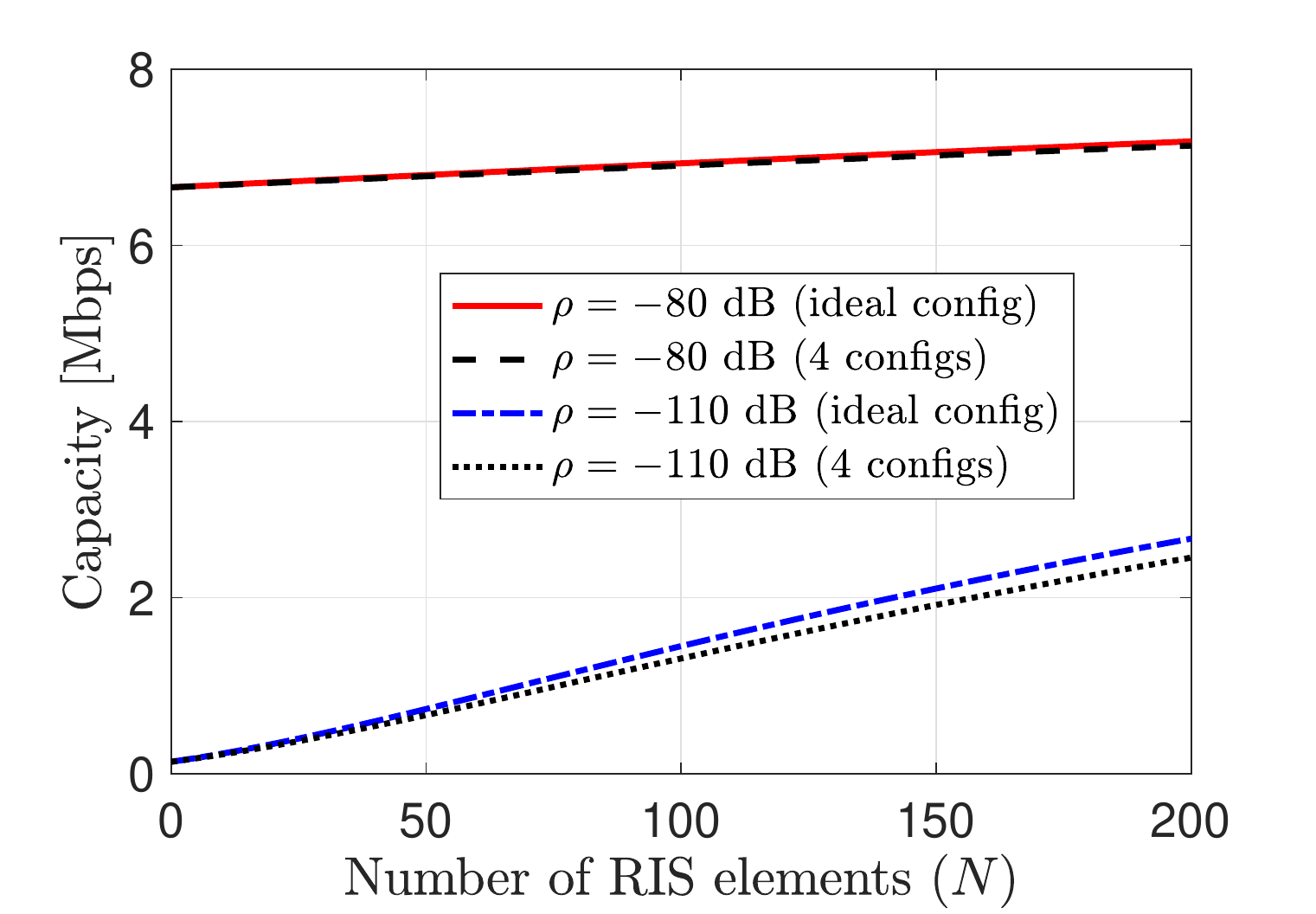}
\end{overpic} \vspace{-2mm}
	\caption{The capacity of a narrowband channel grows with the number of RIS elements. The benefit of controlling the channel using an RIS is particularly large when the uncontrollable ``direct'' channel is weak.}
	\label{fig:narrowbandcommunication} \vspace{-5mm}
\end{figure}

Fig.~\ref{fig:narrowbandcommunication} shows the capacities that can be achieved in a narrowband setup with $B=1$ MHz and a varying number of RIS elements with $\gamma=1$. The propagation losses via the RIS are $\alpha_n=-80$\,dB and $\beta_n=-60$\,dB, while $P/(BN_0)=100$\,dB. We consider two cases for the uncontrollable channel: $-80$\,dB (strong) and $-110$\,dB (weak). We notice that the RIS can increase the capacity by orders-of-magnitude when the uncontrollable channel is weak. The RIS-controlled path is 30\,dB weaker than the uncontrollable path when $N=1$, but since its contribution to the SNR grows as $N^2$, it surpasses the uncontrollable channel in strength when having $N=32$ elements and beyond that, the SNR grows as $N^2$. When the uncontrollable channel is strong, the capacity is already high and the RIS has a limited effect on it because $N=1000$ elements are required before the path via the RIS becomes equally strong. Fig.~\ref{fig:narrowbandcommunication} shows results with the ideal RIS configuration and the case with only four phase-shifts per element. The performance difference is small.

\enlargethispage{3mm}

\vspace{-4mm}

\subsection{Reconfiguration under mobility}

Most wireless channels are time-variant due to user mobility. Yet, many properties of communication systems can be studied using LTI system theory by assuming (approximately) piecewise time-invariant channels, as done so far in this article.
However, the study of Doppler effects due to mobility requires to drop the time-invariance assumption and employ linear time-variant (LTV) system theory.

\begin{figure}
	\centering
	\begin{overpic}[width=\columnwidth,tics=10]{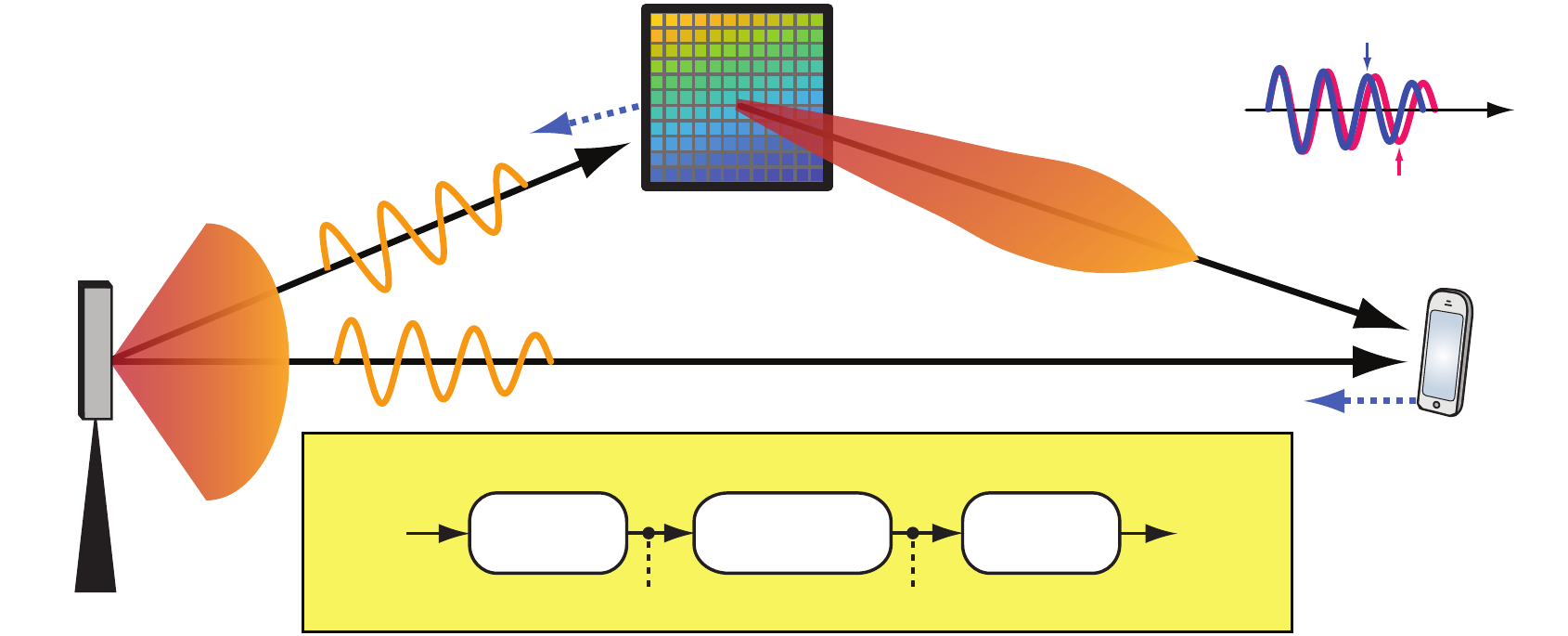}
		\put(1,0){Transmitter}
		\put(81,40){\footnotesize Uncontrollable path }
		\put(81,38){\footnotesize and time-varying RIS}
		\put(86,27.5){\footnotesize Static RIS }
		\put(96,34){\footnotesize Time}
		\put(65,33.4){\footnotesize Received signals:}
		\put(54,38){RIS}
		\put(84,12){User motion}
		\put(22,34){Synthesized}
		\put(22,31.5){RIS motion}
		\put(28,10.5){End-to-end propagation via the $n$th RIS element}
		\put(31.2,6){$a_{n,\pb}(\tau)$}
		\put(44.7,6){$\vartheta_{n,\pb; \theta_n}(\tau, t)$}
		\put(62,6){$b_{n,\pb}(\tau, t)$}
		\put (20,6) {$x_{\pb}(t)$}
		\put (39.7,1.5) {$\tilde x_{n,\pb(t)}$}
		\put (56.6,1.5) {$\bar{x}_{n,\pb(t)}$}
		\put (75,6) {$y_{n,\pb}(t)$}
\end{overpic} \vspace{-7mm}
	\caption{User motion creates a time-variant system, resulting in Doppler shifts. The shifts are generally different for the uncontrollable channel and the controllable channel via the RIS. However, the controllable channel can be configured to achieve a zero Doppler spread by synthesizing movement of the RIS along a matching trajectory.}
	\label{fig:ltv-cascade2} \vspace{-5mm}
\end{figure}

The non-linear Doppler effect widens the signal bandwidth and can result in intersymbol interference, but we will show that the RIS can mitigate some of these effects by varying its configuration to electronically synthesize that it also moves, as illustrated in Fig.~\ref{fig:ltv-cascade2}.
To show how to do that, we start by revisiting the passband input-output relationship for an arbitrary system in \eqref{eq:input-output-passband}, which is given by the convolution equation for LTV filters \cite{Gallager2008}:
\begin{equation} \label{eq:ltv-input-output-passband}
	y_{\pb}(t) = \int_{-\infty}^{\infty} h_{\pb}(u, t) x_{\pb}(t-u) du
\end{equation}
where $h_{\pb}(\tau, t)$ is the real-valued time-varying impulse response which can be regarded as a conventional LTI channel impulse response in $\tau$ that is slowly varying with the time $t$. While \eqref{eq:ltv-input-output-passband} is very similar to the convolution equation for LTI systems in \eqref{eq:input-output-passband}, there is no direct correspondence to $Y_\pb(f)$ in \eqref{eq:input-output-baseband-Fourier1}. Instead, $y_\pb(t)$ is related to the frequency-domain representation of the passband input signal as
\begin{equation} \label{eq:ltv-input-output-freq}
	y_{\pb}(t) = \int_{-\infty}^{\infty} H_{\pb}(f, t) X_\pb(f) e^{j 2 \pi f t} df,
\end{equation}
where $H_{\pb}(f, t)$ is the time-variant transfer function, obtained as the Fourier transform of $h_{\pb}(\tau, t)$ with respect to $\tau$. This function can be regarded as an LTI frequency response that varies slowly with $t$.

Consider the example  system in Fig.~\ref{fig:basic_example}(b) and assume that the receiver is now a mobile user terminal. Then, the uncontrollable channel $h_{d,\pb}(\tau, t)$ and the controllable channels from the RIS elements to the receiver $b_{n,\pb}(\tau, t)$, $n = 1, \dots, N$, are LTV filters. In contrast, the channels $a_{n,\pb}(\tau)$, $n = 1, \dots, N$, are still LTI filters since the transmitter and the RIS are static. To show that an RIS can manage mobility, we need to drop the assumption that the configuration $\vect{\theta}$ is constant and consider each RIS element to also be an LTV filter with time-varying impulse response $\vartheta_{n,\pb; \theta_n}(\tau, t)$ and transfer function $\Theta_{n,\pb; \theta_n}(f, t) = \sqrt{\gamma_{\theta_n}(t)} e^{-j 2 \pi f \tau_{\theta_n}(t)}$, which is analogous to \eqref{eq:ris-element-impulse-response} except that $\gamma_{\theta_n}$ and $\tau_{\theta_n}$ are now functions of the time $t$.

The resulting end-to-end propagation path over the $n$th RIS element is shown as a cascade of three systems in Fig.~\ref{fig:ltv-cascade2}. Computing the joint impulse response for this path is slightly more complicated than before due to the cascade of two LTV filters. To this end, we define the auxiliary signals $\tilde x_{n,\pb}(t)$ and $\bar{x}_{n,\pb}(t)$ as indicated in Fig.~\ref{fig:ltv-cascade2}; that is, as the outputs of the first and second filters, respectively.
From the input-output relations in \eqref{eq:ltv-input-output-passband} and \eqref{eq:ltv-input-output-freq}, it follows that the signal transmitted over the $n$th RIS element is
\begin{align}
y_{n,\pb}(t) &= \int_{-\infty}^\infty b_{n,\pb}(u, t) \bar{x}_{n,\pb}(t - u) du \\
&= \int_{-\infty}^\infty b_{n,\pb}(u, t) \left( \int_{-\infty}^\infty \Theta_{n,\pb; \theta_n}(f, t - u) \tilde X_{n,\pb}(f) e^{j 2\pi f (t - u)} df \right) du \\
&= \int_{-\infty}^\infty \underbrace{\left( \int_{-\infty}^\infty b_{n,\pb}(u, t) \Theta_{n,\pb; \theta_n}(f, t - u) e^{-j 2\pi f u} du \right) A_{n,\pb}(f)}_{=H_{\pb; \theta_n}(f, t)} X_\pb(f) e^{j 2\pi f t}  df. \label{eq:ltv-transfer-func-deriv}
\end{align}
Observe that $\tilde X_{n,\pb}(f)$ exists and is equal to $A_{n,\pb}(f) X_{\pb}(f)$ because the channel from the transmitter to the RIS element is still an LTI system.
	In \eqref{eq:ltv-transfer-func-deriv}, we can identify the joint time-varying transfer function $H_{\pb; \theta_n}(f, t)$ of the $n$th RIS propagation path.
Its corresponding time-variant impulse response $h_{\pb; \theta_n}(\tau, t)$ is obtained from the inverse Fourier transform with respect to $f$. Then, due to the linearity, the time-varying impulse response of the end-to-end system in \eqref{eq:ltv-input-output-passband} is $h_{\pb;\vect{\theta}}(\tau, t) = h_{d,\pb}(\tau, t) + \sum_{n=1}^N h_{\pb; \theta_n}(\tau, t).$

The obtained input-output relation holds for all LTV systems. Let us now assume a narrowband channel
with $A_{n,\pb}(f) = \sqrt{\alpha_n} e^{-j 2 \pi f \tau_{n,a}}$ and $b_{n,\pb}(\tau, t) = \sqrt{\beta_n(t)} \delta(\tau - \tau_{n,b}(t))$. Then, the channel over the $n$th RIS element has the transfer function
\begin{align}
	H_{\pb; \theta_n}(f, t)
	&= \sqrt{\alpha_n  \gamma_{\theta_n}(t) \beta_n(t)} e^{-j 2 \pi f (\tau_{n,a} + \tau_{\theta_n}(t) + \tau_{n,b}(t))}
\end{align}
where we utilized the fact that the RIS element's transfer function can be arbitrarily translated in time. Using this result, the channel output is straightforwardly obtained as
\begin{multline}
	y(t) =
	\sqrt{\rho(t)} e^{-j 2\pi f_c \tau_d(t)} x(t - \tau_d(t)) \\ + \sum_{n=1}^N \sqrt{\alpha_n  \gamma_{\theta_n}(t) \beta_n(t)} e^{-j 2\pi f_c (\tau_{n,a} + \tau_{\theta_n}(t) + \tau_{n,b}(t))} x(t - \tau_{n,a} - \tau_{\theta_n}(t) - \tau_{n,b}(t)).
\end{multline}
This system model allows to study the optimal RIS configuration taking mobility effects into account.

First, consider the case with only a controllable channel; that is, $\rho(t) = 0$ for all $t$.
Recall from \eqref{eq:SNR-vector-description-CS} that the capacity-maximizing configuration in this scenario is 
$\tau_{\theta_n}(t) = \varrho_n(t) - \tau_{n,a}-\tau_{n,b}(t)$
where $\varrho_n(t)$ is chosen such that the signals from the $N$ RIS elements reach the receiver with aligned phases. The solution that achieves this with minimum delay was determined in \eqref{eq:optimized-delays-basic} as $\varrho_n(t) = \max_{i=1,\ldots,N} \{ \tau_{i,a}+\tau_{i,b}(t) \}$ for all $n$.
When this solution is applied under mobility,  the pointwise maximum operation 
will occasionally lead to discontinuities with sudden phase jumps.
Avoiding this requires restricting the phase-shifts induced by $\varrho_n(t)$ to integer multiples of $2\pi$; that is, $\varrho_n(t) = \frac{k_n(t)}{f_c}$ with $k_n(t)$ being a piecewise constant function taking integer values. Moreover, causality requires  all $\tau_{\theta_n}(t)$ to be nonnegative.
Thus, the delay is minimized by 
$k_n(t) = \max_{i=1,\ldots,N} \lceil f_c \left( \tau_{i,a} + \tau_{i,b}(t) \right) \rceil$
for all $n$. This increases the propagation delay by at most one period of the carrier signal compared to the delay-minimizing configuration, avoids undesired spectral effects, and does not introduce any extra delay spread. Moreover, 
we have $x(t - \tau_d(t)) \approx x(t - \tau_{n,a} - \tau_{\theta_n}(t) - \tau_{n,b}(t))$ as long as the symbol time is much larger than the delay spread.

Another phenomenon that only occurs under mobility is  Doppler shift. For each propagation path, the Doppler shift is defined as the 
difference between the observed and emitted frequency:

\begin{equation} \label{eq:doppler-shift}
	\mathcal D_n
	= -\frac{d}{dt} \left[ f_c \left( \tau_{n,a} + \tau_{\theta_n}(t) + \tau_{n,b}(t) \right) \right]
	= -f_c \frac{d \left( \tau_{\theta_n}(t) + \tau_{n,b}(t) \right)}{dt}.
\end{equation}
Interestingly, the Doppler shift can be fully compensated for by the RIS by tuning the delays such that the RIS counteracts the rate of change of $\tau_{n,b}(t)$; that is each RIS element needs to implement $\tau_{\theta_n}(t)$ such that $d \tau_{\theta_n}(t) / dt = - d\tau_{n,b}(t) / dt$.
This technique is known as Doppler cloaking and leads to the Doppler effect not being observable in the received signal. It has been investigated in different contexts to reduce electromagnetic noise caused by moving objects towards radar and sonar systems or to build invisibility cloaks even for moving objects \cite{Ramaccia2017}. Possible applications in communication systems could be to deploy legacy systems in high-mobility scenarios they where not designed for or to connect Internet-of-things devices with very simple transceiver chains to a fast moving satellite in space.

For the SNR-optimal configuration with minimum delay derived earlier, we obtain the Doppler shift
\begin{equation}
	\mathcal D_n = -f_c\left[  \frac{d}{dt} \left( \frac{k_n(t)}{f_c} - \tau_{n,a}-\tau_{n,b}(t) \right) + \frac{d \tau_{n,b}(t)}{dt} \right] = - \frac{d k_n(t)}{dt}.
\end{equation}
Because changes in $k_n(t)$ do not lead to phase discontinuities, it has no effect on the Doppler shift and $\frac{d k_n(t)}{dt}$ can be assumed zero from a practical perspective. Hence, this configuration maximizes the SNR with minimum delay and also removes Doppler shifts \cite{Matthiesen2020a}.

Next, consider the case with an additional uncontrollable propagation path (i.e., $\rho(t) \neq 0$), again under the assumption that it is shorter than the RIS path. Following the previous discussion, the SNR-optimal configuration with minimum delay is
$\tau_{\theta_n}(t) = \left(\tau_d(t) - \tau_{n,a}-\tau_{n,b}(t) \right) + \frac{k_{n,\min}(t)}{f_c}$,
where $k_{n,\min}(t)$ is the minimum integer that satisfies the causality constraint; that is, $k_{n,\min}(t) = \lceil f_c \left( \tau_{n,a} + \tau_{n,b} - \tau_d \right) \rceil$. Using this configuration and employing \eqref{eq:doppler-shift}, the Doppler shift of the propagation path over the $n$th RIS element is
$\mathcal D_n = -f_c \frac{d}{dt} \tau_d(t)$,
where we omitted $\frac{d}{dt} k_{n,\min}(t)$ for the same reason as before. This is the same as the Doppler shift of the uncontrollable channel and, hence, the Doppler spread is zero.

It is impossible to simultaneously maximize the SNR and compensate the Doppler shifts introduced by the RIS (as is the case without an uncontrollable path). However, even if it was possible, it would be undesired as it results in a Doppler spread of $f_c \frac{d}{dt} \tau_d(t)$. As the mitigation of Doppler spreads is usually much more difficult than treating Doppler shifts at the receiver, having the RIS not introducing additional Doppler spread in the system could be considered the optimal solution in terms of Doppler effects.

In conclusion, we have observed that the SNR-optimal configuration with minimum delay obtained using the developed LTI system model is still valid and optimal when mobility is involved. While some care has to be taken not to introduce additional frequency components into the spectrum due to phase discontinuities, an SNR-maximizing configuration also minimizes the delay spread and does not introduce additional Doppler spread into the system.

\subsection{RIS design for wideband capacity maximization}

The RIS optimization becomes more challenging in the wideband case where there are $K$ parallel subcarriers, each represented by the system model in \eqref{eq:symbol-sampled-model-OFDM}. The subcarriers are separate AWGN channels but share the power $P$ since they are transmitted simultaneously.
Suppose the power $P_{\nu}=\mathbb{E}\{ | \bar{x}[\nu] |^2 \}$ is assigned to subcarrier $\nu$. Any power allocation $P_0,\ldots,P_{K-1}$ satisfying $P= \frac{1}{K} \sum_{\nu=0}^{K-1} P_{\nu}$ is feasible.

When studying this setup, we make use of the frequency response in \eqref{eq:OFDM-response} for the controllable channel via the RIS but we also add an uncontrollable channel. We let $\vect{h}_d = [h_d[0], \ldots, h_d[M-1]]^{\Ttran}$ denote the discrete-time impulse response of the uncontrollable channel, which can be computed similar to \eqref{eq:example-end-to-end}:
$h_d[k] = \sum_{l=1}^{L_d} \sqrt{\rho^{l}} e^{-\imagunit 2\pi f_c \tau_{d}^{l}}
\sinc ( k + B( \eta -\tau_{d}^{l}) )$ where $L_d$ is the number of paths, $\rho^{l} \in [0,1]$ is propagation loss of the $l$th path, and $\tau_{d}^{l} \geq 0$ is its delay.

For a given RIS configuration $\vect{\theta}$ and power allocation, the so-called achievable rate is 
\begin{align} \label{eq:rate-OFDM}
    R = \frac{B}{K+M-1} \sum_{\nu=0}^{K-1} \log_2 \left( 1 + \frac{P_{\nu} | \vect{f}_{\nu}^{\Htran} \vect{h}_d + \vect{f}_{\nu}^{\Htran} \vect{V}^{\Ttran} \vect{\omega}_{\vect{\theta}}|^2}{B N_0} \right) \quad \textrm{bit/s}
\end{align} 
where $\vect{f}_{\nu}^{\Htran}$ is the $\nu$th row of the DFT matrix $\vect{F}$. This rate expression is a summation over the $K$ subcarriers, which is then divided by $K+M-1$ (instead of $K$) to compensate for the cyclic prefix loss. The capacity is obtained by maximizing this expression with respect to both the power allocation and RIS configuration. The former is a classical problem with a solution called waterfilling power allocation \cite{Yang2020a}:
\begin{equation} \label{eq:waterfilling}
    P_{\nu} = \max \left( \mu - \frac{B N_0}{| \vect{f}_{\nu}^{\Htran} \vect{h}_d + \vect{f}_{\nu}^{\Htran} \vect{V}^{\Ttran} \vect{\omega}_{\vect{\theta}}|^2}, 0\right)
\end{equation}
where the parameter $\mu \geq 0$ is selected to make $\frac{1}{K} \sum_{\nu=0}^{K-1} P_{\nu}=P$.

The maximization of \eqref{eq:rate-OFDM} with respect to the RIS configuration $\vect{\theta}$ entails selecting the most preferred vector $\vect{\omega}_{\vect{\theta}}$ among those that the hardware can generate. Note that the same vector affects all subcarriers because the transmissions are simultaneous. In the narrowband case, we could optimize the RIS in closed form since there was only one channel (one subcarrier), but now we need to find a nontrivial tradeoff between all $K$ subcarriers. 
So far, this problem seems mathematically intractable to solve to global optimality, thus the literature contains heuristic solutions based on successive convex approximation, semidefinite relaxation, and strongest tap maximization (STM) in the time domain \cite{Zheng2020,Yang2020a,Lin2020a}. In this article, we will focus on the STM solution from \cite{Zheng2020,Lin2020a} and compare it against an upper bound.

The intuition behind STM is that the received signal power is spread out over the $K$ subcarriers but rather concentrated in the time domain since $M \ll K$ \cite{Zheng2020}. Hence, selecting a configuration $\boldsymbol{\theta}$ that is good for one strong channel tap is better than an arrangement that is good for one strong subcarrier. This is particularly true when there is an LOS propagation path that is much stronger than all other paths.
When adding the uncontrollable channel to \eqref{eq:example-end-to-end-vectors}, the $\ell$th tap of the impulse response becomes $h_d[\ell] + \vect{v}_{\ell}^{\Ttran} \vect{\omega}_{\vect{\theta}}$. We begin by finding the value of $\vect{\omega}_{\vect{\theta}}$ that maximizes the magnitude of each tap:
\begin{equation} \label{eq:STM-step1}
    \vect{\omega}_{\ell} = \argmax{\vect{\omega}_{\vect{\theta}}} \,\, | h_d[\ell] + \vect{v}_{\ell}^{\Ttran} \vect{\omega}_{\vect{\theta}} |, \quad  \ell = 0,\ldots,M-1.
\end{equation}
In STM, we then select the one of candidate solutions $\vect{\omega}_{0},\ldots,\vect{\omega}_{M-1}$ resulting in the largest magnitude:
\begin{equation}
    \vect{\omega}_{\textrm{STM}} = \vect{\omega}_{\ell^{\textrm{opt}}} \quad \textrm{where } \,  \ell^{\textrm{opt}} = \argmax{\ell\in \{ 0, \ldots, M-1\}}
| h_d[\ell] + \vect{v}_{\ell}^{\Ttran} \vect{\omega}_{\ell} |.
\end{equation}
Each of the subproblems in \eqref{eq:STM-step1} can be solved analogously with the narrowband SNR maximization in \eqref{eq:SNR-max-direct-channel}.
The solution is $\vect{\omega}_{\ell}= [e^{\imagunit (\arg(h_d[\ell])- \arg([\vect{v}_{\ell}]_1)) }, \ldots, e^{\imagunit (\arg(h_d[\ell])- \arg([\vect{v}_{\ell}]_N) )}]^{\Ttran}$, where $[\vect{v}_{\ell}]_n$ denotes the $n$th entry of $\vect{v}_{\ell}$ and $\arg(\cdot)$ gives the argument (phase) of a complex number. Note that this solution rotates the phase of each term in the inner product $\vect{v}_{\ell}^{\Ttran} \vect{\omega}_{\ell}$ so it matches with the phase of $h_d[\ell]$.

To evaluate the quality of the heuristic STM solution, we can compare the rate that it achieves with an upper bound. Suppose we could select a different value of $\vect{\omega}_{\vect{\theta}}$ on each subcarrier. We could then jointly maximize the SNRs of all subcarriers.
For the $\nu$th subcarrier, its SNR is maximized by selecting
$\vect{\omega}_{\vect{\theta}}= [e^{\imagunit (\arg(\vect{f}_{\nu}^{\Htran} \vect{h}_d)- \arg([\vect{f}_{\nu}^{\Htran} \vect{V}^{\Ttran}]_1)) }, \ldots, e^{\imagunit (\arg(\vect{f}_{\nu}^{\Htran} \vect{h}_d)- \arg([\vect{f}_{\nu}^{\Htran} \vect{V}^{\Ttran}]_N)) }]^{\Ttran}$.
The resulting upper bound is
\begin{align} \label{eq:rate-OFDM-upper-bound}
    R \leq \frac{B}{K+M-1} \sum_{\nu=0}^{K-1} \log_2 \left( 1 + \frac{P_{\nu} }{B N_0} \left( | \vect{f}_{\nu}^{\Htran} \vect{h}_d| + \|\vect{f}_{\nu}^{\Htran} \vect{V}^{\Ttran} \|_1 \right)^2  \right)
\end{align} 
where $\| \cdot \|_1$ denotes the $L_1$ norm. This upper bound is only exactly achievable in the unlikely event that the same RIS configuration happens to maximize the SNRs of all subcarriers.

\begin{figure}[t!]
        \centering
        \begin{subfigure}[b]{.47\columnwidth} \centering  
	\begin{overpic}[width=\columnwidth,tics=10]{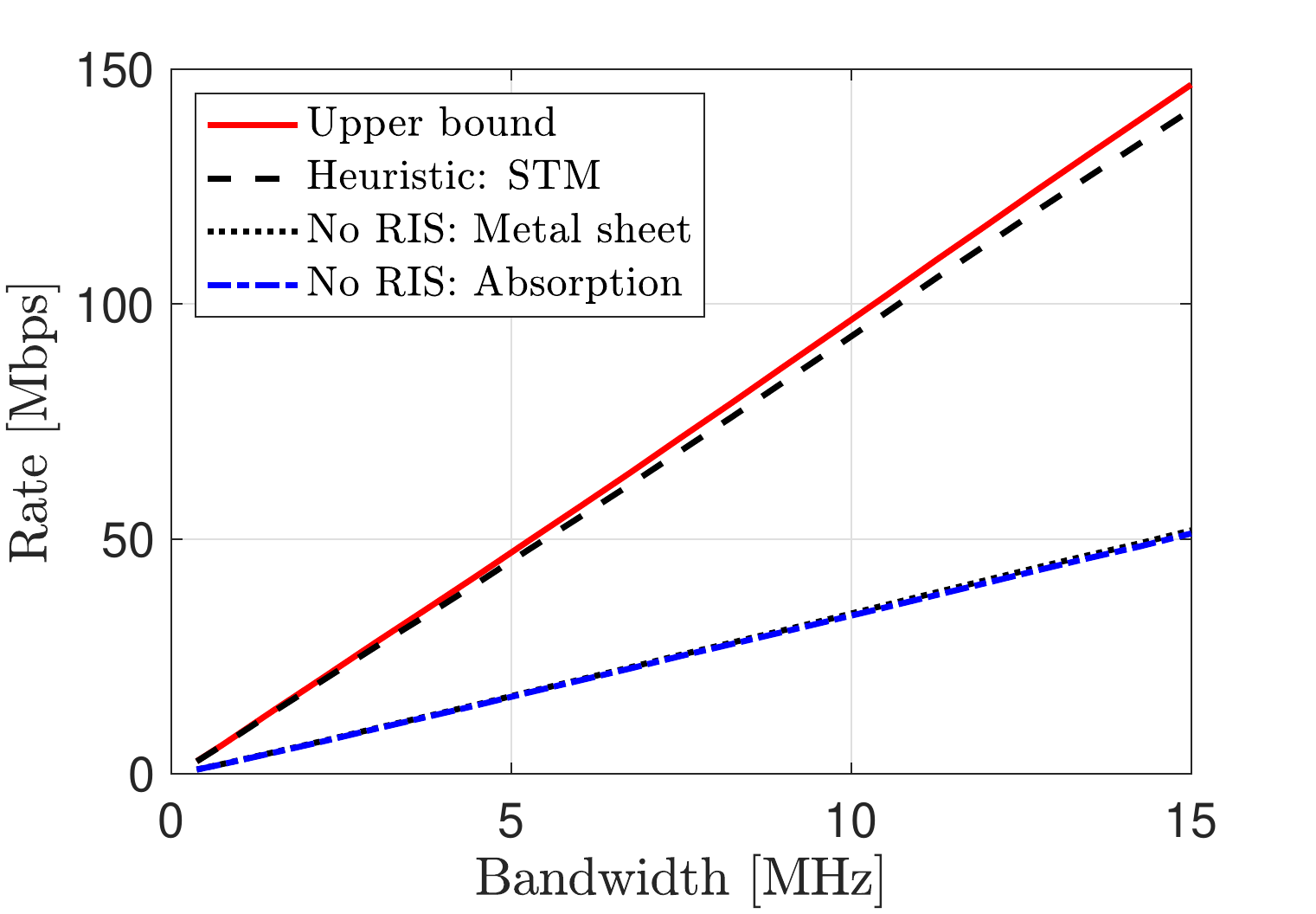}
\end{overpic} 
                \caption{LOS channel from transmitter to RIS.} 
                \label{fig:OFDMcommunication:LOS}
        \end{subfigure} 
        \begin{subfigure}[b]{.47\columnwidth} \centering 
	\begin{overpic}[width=\columnwidth,tics=10]{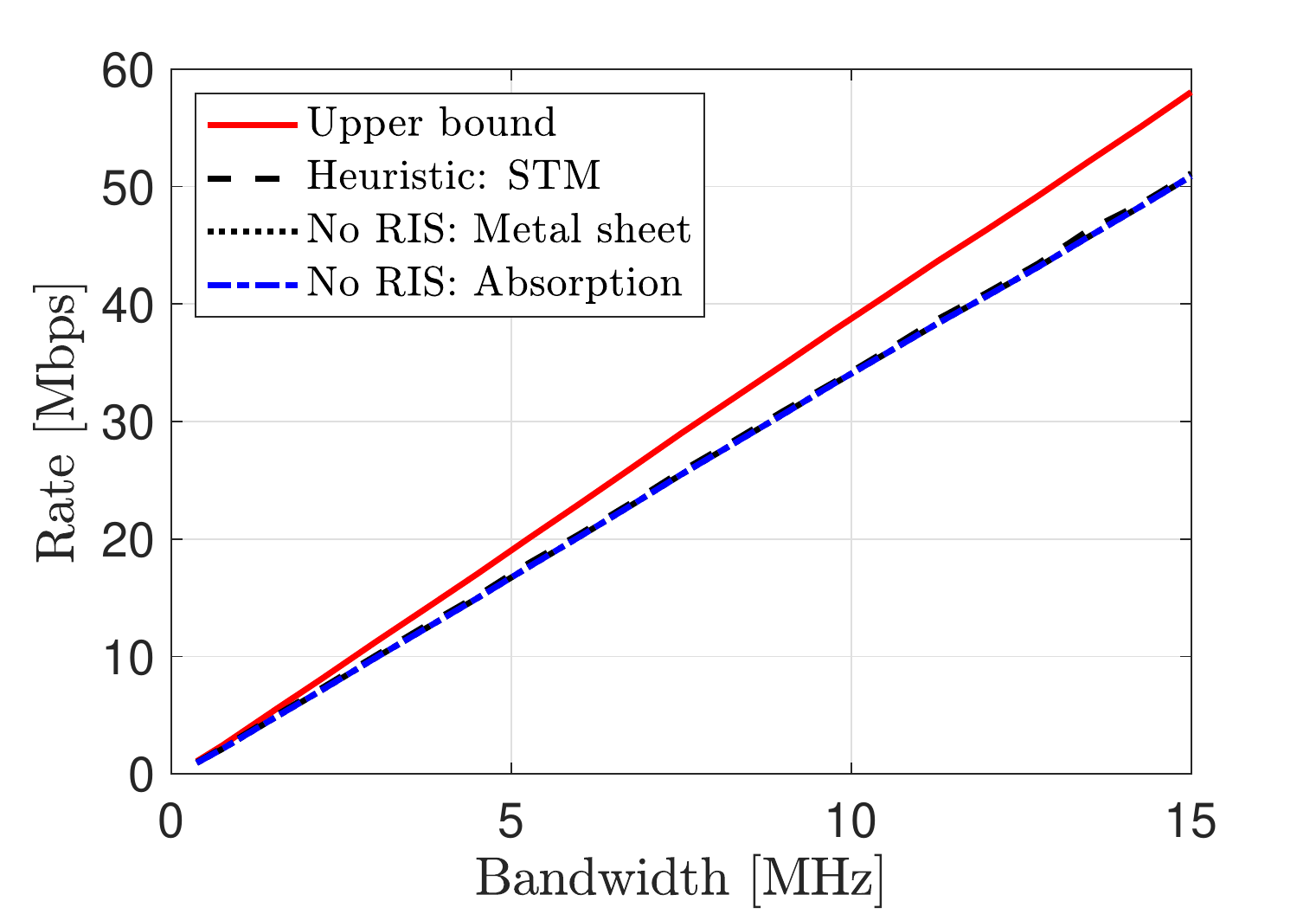}
\end{overpic}
                \caption{NLOS channel from transmitter to RIS.} 
                \label{fig:OFDMcommunication:NLOS}
        \end{subfigure} \vspace{-3mm}
        \caption{The rate that can be achieved over a wideband channel grows proportionally to the bandwidth. If there are LOS channels to and from the RIS, it can greatly improve the slope and achieve a rate close to the upper bound in \eqref{eq:rate-OFDM-upper-bound}. This performance gain collapses if there is an NLOS path to the RIS.}
        \label{fig:OFDMcommunication}
\end{figure}

Fig.~\ref{fig:OFDMcommunication} shows simulation results for the achievable rates over a wideband channel, inspired by the setup in Fig.~\ref{fig:basic_example}(b).
The RIS and receiver are located in a large room and, thus, have an LOS channel between them. The transmitter is an access point located 400 meters away and has an NLOS  channel to the receiver. We will consider cases where the transmitter-to-RIS channel is either LOS or NLOS. The carrier frequency is 3 GHz and the RIS is $0.5 \times 0.5$ m, which corresponds to $N=400$ elements that each have dimension $\lambda/4 \times \lambda/4$.
The channels are modeled similar to the 3GPP channel model in \cite{3GPP25996} and the rate is averaged over random realizations of the multipath components.
The rate in \eqref{eq:rate-OFDM} is shown in Fig.~\ref{fig:OFDMcommunication} as a function of the bandwidth $B$. The optimal waterfilling power allocation from \eqref{eq:waterfilling} is utilized and the transmit power $P$ grows proportionally to the bandwidth. The subcarrier spacing is 150 kHz, thus the number of subcarriers increases with $B$ as well as the number of channel taps.

Fig.~\ref{fig:OFDMcommunication}(a) considers the case with an LOS path from the transmitter to the RIS.
The dashed curve represents the rate when using the heuristic STM configuration of the RIS. It provides 96-98\% of the upper bound from \eqref{eq:rate-OFDM-upper-bound}. The gap grows with $B$ due to the increased frequency-selectivity, but since the LOS paths to/from the RIS are stronger than the scattered paths, it is possible to find a single RIS configuration that works well over the entire band. The refined RIS configuration algorithms described in \cite{Zheng2020,Yang2020a,Lin2020a} can reduce the gap but only improve the rate by a few percent. 
It is interesting to compare the rate with what could be achieved without an RIS. In this case, we can either replace the RIS with an absorbing material, thereby removing all the paths via the RIS, or by a passive metal sheet causing zero phase-shifts. The corresponding curves in Fig.~\ref{fig:OFDMcommunication}(a) are nearly overlapping but there are ideal situations where a perfectly rotated metal sheet is almost as efficient as an RIS \cite{Ozdogan2019a}.
The RIS can increase the rate by 2.7-2.9 times, which makes a huge difference when there are several MHz of bandwidth.

Fig.~\ref{fig:OFDMcommunication}(b) considers the case with an NLOS path to the RIS, which has two effects: The path via the RIS is weaker and there is no dominant path. The former effect results in a much smaller gap between the upper bound and ``no RIS'' cases, while the latter  results in an inability to find a single RIS configuration that fits the entire band. In this case, the RIS can improve the rate by 4\% in the narrowband case of $B=400$\,kHz but the gain vanishes as $B$ increases.
One can find a slightly better RIS configuration using the algorithms in \cite{Zheng2020,Yang2020a,Lin2020a}, but the bottomline is that an RIS must be carefully deployed to be truly effective. It should be deployed, as in Fig.~\ref{fig:OFDMcommunication}(a), at a location with LOS to the access point and can then be configured to greatly improve the rate to users that are within the LOS of it.

\subsection{Protocol for channel estimation and reconfiguration}

The capacity maximization has been described above assuming that the channels are perfectly known, but a preceding channel estimation phase is required in practice. 
Since the RIS is passive, the estimation must be carried out at the receiver.
If we focus on the OFDM case and neglect the uncontrollable channel, the frequency response in \eqref{eq:OFDM-response} is $\bar{\vect{h}}_{\vect{\theta}}= \vect{F} \vect{V}^{\Ttran} \vect{\omega}_{\vect{\theta}}$, where 
the matrix $\vect{V}$ represents the cascade of the channel from the transmitter to the RIS and the channel from the RIS to the receiver.
It is sufficient to estimate $\vect{V}$ to compute $\bar{\vect{h}}_{\vect{\theta}}$ for any $\vect{\omega}_{\vect{\theta}}$, which is fortunate since it is hard to resolve the individual channels.

Suppose a known pilot signal $x$ is transmitted on each of the $M$ subcarriers, where $M$ equals the number of (unknown) terms in the time-domain channel $\vect{h}_{\vect{\theta}}$. Let the RIS configuration be $\vect{\theta}_t$, where $t$ is the index of the OFDM block. The received signal $\bar{\vect{z}}[t] \in \mathbb{C}^M$ over the $M$ pilot-bearing subcarriers at OFDM block $t$ is 
\begin{equation}  \label{eq:channel-estimation1}
    \bar{\vect{z}}[t] = \vect{F}_M \vect{V}^{\Ttran} \vect{\omega}_{\vect{\theta}_t} x + \bar{\vect{w}}[t]
\end{equation}
where $\vect{F}_M$ contains the $M$ rows of $\vect{F}$ corresponding to the selected subcarriers and $\bar{\vect{w}}[t] \in \mathbb{C}^M$ contains the corresponding noise. There are $MN$ unknown parameters in $\vect{V}$ but we only obtain $M$ observations from $\bar{\vect{z}}[t]$. Using more than $M$ subcarriers for pilot transmission will not resolve this issue since the impact of the RIS configuration  $\vect{\omega}_{\vect{\theta}_t}$ cannot be removed from \eqref{eq:channel-estimation1}; a vector is non-invertible.
The way to get $MN$ linearly independent observations, where $N$ is the number of RIS elements, is to consider a sequence of $N$ OFDM blocks  
with different configurations: $\vect{\theta}_1,\ldots,\vect{\theta}_N$. The joint received signal is
\begin{equation}  \label{eq:channel-estimation2}
    \underbrace{\big[ \bar{\vect{z}}[1], \ldots, \bar{\vect{z}}[N] \big]}_{=\bar{\vect{Z}}} = \vect{F}_M \vect{V}^{\Ttran} \underbrace{[\vect{\omega}_{\vect{\theta}_1}, \ldots , \vect{\omega}_{\vect{\theta}_N}]}_{=\vect{\Omega}} x + \underbrace{\big[ \bar{\vect{w}}[1] , \ldots , \bar{\vect{w}}[N] \big]}_{=\bar{\vect{W}}}.
\end{equation}
If the RIS configurations are selected so that $\vect{\Omega}$ is invertible (e.g., a DFT matrix), we can rewrite 
\eqref{eq:channel-estimation2} as
\begin{equation}  \label{eq:channel-estimation3}
\underbrace{\frac{1}{x}\vect{F}_M^{-1}\bar{\vect{Z}} \vect{\Omega}^{-1}}_{\textrm{Known signal}} = \vect{V}^{\Ttran}  + \underbrace{\frac{1}{x}\vect{F}_M^{-1} \bar{\vect{W}} \vect{\Omega}^{-1}}_{\textrm{Noise}}.
\end{equation}
This is a linear model from which a variety of classical channel estimation techniques can be applied. In fact, \eqref{eq:channel-estimation3} is already the least-square estimate of $\vect{V}^{\Ttran}$. 
If there is prior information, such as a fading distribution or spatial-temporal sparsity, this can be used to devise better estimators that also require shorter pilots \cite{Wu2019a,Bjornson2020a,Taha2021a}.
Since switching between configurations is a non-linear operation and, thus, can modulate the reflected signals into other bands, it should be done in a silent guard interval in between OFDM blocks.
After the receiver has estimated the channel, it can compute a suitable configuration $\vect{\omega}_{\vect{\theta}}$ (as described earlier) that can be utilized as long as the channel remains static. The control channel described in Fig.~\ref{fig:LayeredModel} can be utilized to inform the RIS of the desired configuration.
Since only $M$ out of $K$ subcarriers are used for pilots, the remaining ones can carry data. To handle mobility, one can develop protocols for progressive RIS reconfiguration where data is continuously transmitted 
and pilots are sent at regular intervals to re-estimate the channel and reconfigure the RIS \cite{Zheng2020,Lin2020a}.

\section{RIS-aided localization and sensing}

We will now consider localization and sensing.
The objective of localization is to estimate and track the location of an actively communicating user device, while the objective of sensing is to estimate and track the location of passive objects or users. All radio localization and sensing systems operate under common principles:
there are location references, dynamic user states, and measurements, which are connected to the user state via a statistical model. 
The development of a radio localization
system has three main components: design, channel estimation, and localization/sensing. 
\emph{Design} includes the placement and configuration of {reference points},
and the design of pilot signals to maximize localization accuracy.
The design can be offline, but also online, to adapt to current
user location and requirements. An important tool in the design phase is Fisher information theory \cite{shen2010fundamental}. 
\emph{Channel parameter estimation} is usually performed prior to localization and sensing, and involves estimation of geometric parameters (e.g., delays,
angles, frequency shifts) from received signals.  Note that both localization and
communication rely on channel knowledge. However, localization explicitly determines the geometric parameters, while the unstructured channel  \eqref{eq:channel-estimation3} is sufficient for communication. 
\emph{Location estimation, sensing, and tracking} are performed after channel estimation, with the aim to invert the geometric relation between the user's location and the channel
parameter estimates to recover the user's location as well as the state of passive objects.
Tracking algorithms (e.g., the extended Kalman filter) are used to recursively update these locations over time. Localization and sensing most often involve determination
of nuisance parameters (e.g., synchronization and other biases, as well as data associations between measurements and objects), leading to high-dimensional and nonlinear optimization problems. 
While a detailed treatment of localization and tracking techniques is beyond the scope of this tutorial article, we provide a brief overview of localization and sensing without RIS in 4G and 5G, to contrast with the potential benefits that an RIS brings.

\vspace{-4mm}

\subsection{Localization and sensing in 4G and 5G}

\begin{figure}
\begin{center} \vspace{-4mm}
	\begin{overpic}[width=.9\columnwidth,tics=10]{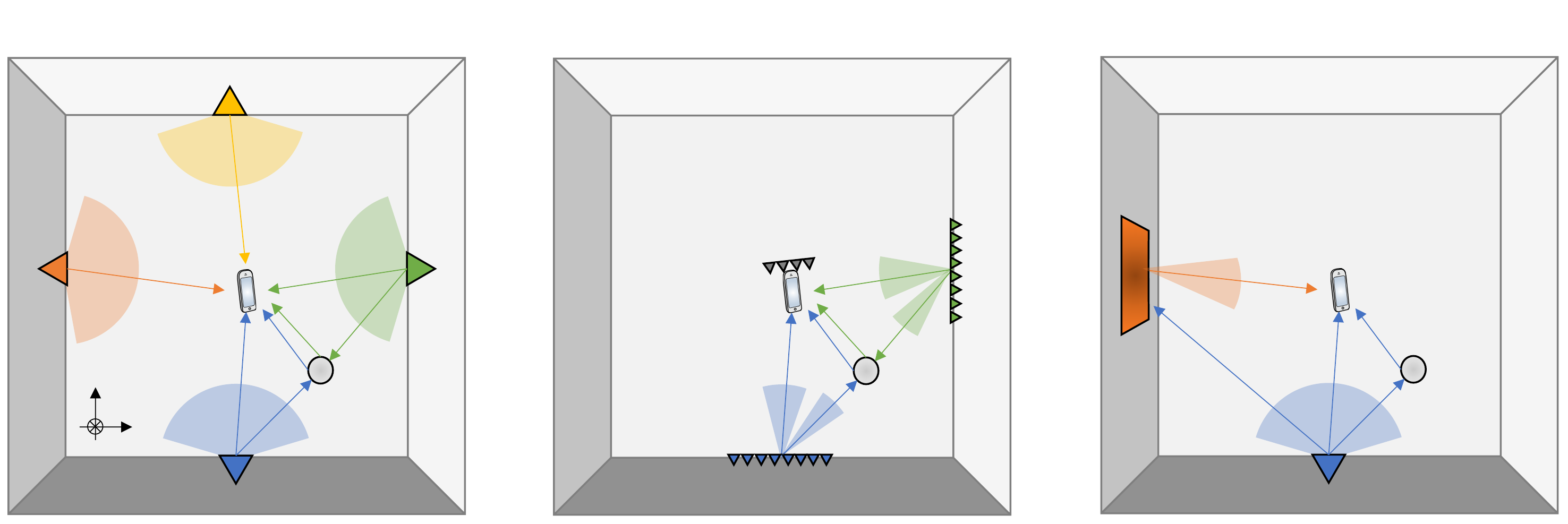}
		\put(10,31){\footnotesize 4G: TDOA}
		\put(39,31){\footnotesize 5G: TDOA+AOA+AOD}
		\put(73.5,31){\footnotesize Beyond 5G: TDOA+AOD}
		\put(11.5,2){\footnotesize BS}
		\put(49,2){\footnotesize BS}
		\put(81,2){\footnotesize BS}
		\put(1,14){\footnotesize BS}
		\put(27,14){\footnotesize BS}
		\put(61.5,16){\footnotesize BS}
		\put(16,26.5){\footnotesize BS}
		\put(11,13){\footnotesize User}
		\put(45.2,14){\footnotesize User}
		\put(80.5,13){\footnotesize User}
		\put(21.5,8){\footnotesize SP}
		\put(56.2,8){\footnotesize SP}
		\put(91,8){\footnotesize SP}
		\put(70.4,20){\footnotesize RIS}
		\put(86.5,15){\footnotesize $\posUE$}
		\put(85.5,2){\footnotesize $\pos_{\text{BS}}$}
		\put(91,11.5){\footnotesize $\pos_{\text{SP}}$}
		\put(70.2,10){\footnotesize $\pos_{\text{RIS}}$}
\end{overpic} \vspace{-4mm}
\end{center} 
\caption{\label{fig:Localization-and-sensing} Localization and sensing across
generations of mobile communication technology: 4G relies on multiple BSs to obtain TDOA measurements.
Uncontrollable multipath components, represented here by an SP, become a disturbance.
In 5G, the inclusion of AOA and AOD measurements reduces the infrastructure
needs and allows sensing of the environment. Beyond 5G, a scenario with a single BS and an RIS is shown. This infrastructure is sufficient to localize a user and provide partial map information, if we exploit that the RIS path is controllable.}   \vspace{-4mm}
\end{figure}

Each new generation of mobile communications introduces new features for higher-rate communication that also enable more accurate
localization \cite{del2017survey}, as visualized in Fig.~\ref{fig:Localization-and-sensing}. 
 In 4G systems, localization is based on the transmissions of 
pilot signals, sent by multiple synchronized
base stations (BSs) over orthogonal subcarriers. The pilot design is such
that it covers the entire signal bandwidth and avoids inter-BS interference. 
The user estimates the time-of-arrival (TOA) with respect to each BS, which depends on the distance to the BS and the user's clock bias. Estimating TOAs from at least four BSs in LOS allows the user to compute three time-difference-of-arrival (TDOA) measurements, and solve for its 3D location.
The estimation accuracy
depends on the SNR as well as the bandwidth spanned by the pilot signals, which determines the sampling rate and thereby the resolvability of the multipath components in time. In fact, multipath
limits the accuracy to tens of meters in 4G \cite{del2017survey}. In systems with a large bandwidth, the individual multipath components can be resolved and related to physical objects (e.g., a scatter point (SP)) in the environment \cite{leitinger2019belief}. 

In 5G systems operating in mmWave bands, the BS and possibly the user are equipped with multiple antennas \cite{buehrer2018collaborative}. The channel is then parameterized by both delays (as in 4G) and angles: angles-of-arrival (AOA) at the receiver and angles-of-departure (AOD) at the transmitter, both  in azimuth and elevation.  This means that the user can be localized from the AOD of two BSs (by the intersection of two lines), significantly reducing infrastructure needs. The channel parameter measurements can be related to objects in the environment with unknown 3D locations through simultaneous localization and mapping (SLAM). In contrast to 4G, which must collect  measurements over time \cite{leitinger2019belief}, the additional angle  measurements in 5G enable sensing of the environment from a single snapshot
of observations \cite{ge20205g,Witrisal2016}.
Despite the high interest in mmWave bands in 5G, it is important to note that lower frequency bands remain relevant due to their large coverage and support for spatial multiplexing of many users. For dense multipath environment, it is challenging to resolve individual propagation paths, limiting the use of lower bands to favorable propagation environments (e.g., outdoors) or requiring data-driven fingerprinting techniques.

 \vspace{-2mm}

\subsection{Localization and sensing with an RIS}

The inclusion of an RIS provides several new opportunities for localization
\cite{bourdoux20206g,Wymeersch2020b}. They are new synchronized location references and configurable for optimizing localization performance. Each RIS also leads to several new geometric measurements, which in turn improves localization accuracy and coverage. 
The 5G scenario in Fig.~\ref{fig:Localization-and-sensing} is easily generalized to a scenario with an RIS \cite{elzanaty2020reconfigurable}, making the problem highly over-determined and in a sense easier. Thus, 
we will focus on the more refined and challenging case with one single-antenna transmitting BS, one single-antenna receiving user, and one RIS with $N$ elements.
While communication uses approximately sinc-shaped pulses that are bandwidth-efficient, localization uses approximately square-shaped  pulses because these are easier to distinguish in time.

Suppose the uncontrollable channel from the transmitter (i.e., BS) to the receiver (i.e., UE) consists of $L_d$ propagation paths, where $\rho^{l} \in [0,1]$ is the propagation loss and $ \tau_d^{l}\geq 0$ is the delay of the $l$th path. The first one is the LOS path.
Furthermore, the RIS is in the far-field and there is only an LOS path to/from the RIS (i.e., $L_a = L_b = 1$) where $\tau_{\text{RIS}}$ denotes the propagation delay via the first element. Under these circumstances, if $\bar{\vect{x}} \in \mathbb{C}^K$ is the transmitted pilot signal over the $K$ subcarriers, the received signal $\bar{\vect{z}}[t]$ in the OFDM block with index $t$ is

\begin{equation} \label{eq:symbol-sampled-model-OFDM-Z10}
\bar{\vect{z}}[t] = \underbrace{\sum_{l=1}^{L_d} g^l \vect{d} (\tau_d^{l}) \odot \bar{\vect{x}}}_{\text{Uncontrollable channel}} + \underbrace{\vect{v}^{\Ttran}\vect{\omega}_{\vect{\theta}}  \left(\vect{d} (\tau_{\text{RIS}}) \odot \bar{\vect{x}}\right)}_{\text{Controllable channel}} + \bar{\vect{w}}[t]
\end{equation}
where $\odot$ is the Hadamard product, $g^l = \sqrt{\rho^{l}} e^{-\imagunit 2\pi f_c \tau_d^{l}}$ is the complex channel gain of path $l$. Here $\vect{d} (\tau) \in \mathbb{C}^K$ with $\left[\vect{d} (\tau) \right]_{\nu} =  e^{-\imagunit 2 \pi \tau \Delta_f \nu}$ where $\Delta_f = B/K$ is the subcarrier spacing. The vectors $\vect{v}  \in \mathbb{C}^N$ and $\vect{\omega}_{\vect{\theta}}  \in \mathbb{C}^N$ are defined in~\eqref{eq:example-end-to-end-vectors}.
Due to the far-field LOS assumption, we have $\alpha_n = \alpha $ and $ \beta_n = \beta$, for all $n$, and we can make use of the steering vector of the RIS. Recall that we call it $\vect{a}(\vect{\phi})\in \mathbb{C}^N$ where $\vect{\phi}= [{\phi}^{\text{az}}, {\phi}^{\text{el}}]^{\Ttran}\in\mathbb{R}^2$ contains the azimuth and elevation angle. The steering vector gives the phase-shifts between the RIS elements for a plane wave impinging from $\vect{\phi}$.
If we further assume that the fraction of re-radiated power from all RIS elements is identical, i.e., $\gamma_{\theta_n}=\gamma$ in $\vect{\omega}_{\vect{\theta}}$, we can write $\vect{v}^{\Ttran}\vect{\omega}_{\vect{\theta}}$ in~\eqref{eq:symbol-sampled-model-OFDM-Z10} as
\begin{align}\label{eq:vector_b}
    \vect{v}^{\Ttran}\vect{\omega}_{\vect{\theta}} =  g_{\text{RIS}}\bigg(\underbrace{\vect{a}(\vect{\phi}_{a})\odot\vect{a}(\vect{\phi}_{b})}_{=\vect{b}(\vect{\phi}_{b})}\bigg)^{\Ttran}{\bar{\vect{\omega}}_{\vect{\theta}}}
\end{align}
where $g_{\text{RIS}} = \sqrt{\alpha \beta \gamma} e^{j \psi_{\text{RIS}}}$ with $\psi_{\text{RIS}}$ 
being a global phase-shift 
and $\vect{\phi}_{a} =[\phi_{\text{a}}^{\text{az}} , \phi_{\text{a}}^{\text{el}}]^{\Ttran} \in\mathbb{R}^2$ and $\vect{\phi}_{b}=[\phi_{\text{b}}^{\text{az}} , \phi_{\text{b}}^{\text{el}}]^{\Ttran} \in\mathbb{R}^2$ are the (azimuth and elevation) AOA and AOD at the RIS, respectively. The vector ${\bar{\vect{\omega}}_{\vect{\theta}}}  \in \mathbb{C}^N$ is obtained from $\vect{\omega}_{\vect{\theta}}$ by setting $\gamma_n=1$ for all $n$ and, thus, has entries on the unit circle. If the geographical locations $\pos_{\text{BS}}\in \mathbb{R}^3$ and $\pos_{\text{RIS}} \in \mathbb{R}^3$ of the BS and RIS, respectively, are known, so is the AOA $\vect{\phi}_{a}$ and we use the notation $\vect{b}(\vect{\phi}_{b})=\vect{a}(\vect{\phi}_{a})\odot\vect{a}(\vect{\phi}_{b})$ in \eqref{eq:vector_b} to focus on the unknown angle $\vect{\phi}_{b}$.

In the system model above, there are
$3L_d+5$ real unknown parameters: $L_d+1$ complex channel gains $\{g_{\text{RIS}}, g^l: l=1,\ldots,L_d\}$, $L_d+1$ delays $\{\tau_{\text{RIS}}, \tau_d^l: l=1,\ldots,L_d\}$, and a two-dimensional AOD vector $\vect{\phi}_{b}$. 
Let $\posUE\in \mathbb{R}^3$ denote the unknown user location and $\pos_{\text{SP},l}\in \mathbb{R}^3$ denote the location of the $l$th SP, for $l=2,\ldots,L_d$.
These location parameters are related to the system model parameters as follows: 
\begin{align}
\tau_d^1 & = \|\pos_{\text{BS}}-\posUE\|/c+\Delta_{\mathrm{clk}},\label{eq:HWtau0}\\
\tau_d^l & =\|\pos_{\text{BS}}-\pos_{\text{SP},l}\|/c+\|\pos_{\text{SP},l}-\posUE\|/c+\Delta_{\mathrm{clk}}, \quad l>1,\label{eq:HWtaul}\\
\tau_{\text{RIS}} & =\|\pos_{\text{BS}}-\pos_{\text{RIS}}\|/c+\|\posUE-\pos_{\text{RIS}}\|/c+\Delta_{\mathrm{clk}},\label{eq:HWtaur}\\
\phi_{\text{b}}^{\text{az}} & =\arctan \! 2([\vect{R}^{\Ttran}(\posUE-\pos_{\text{RIS}})]_{2},[\vect{R}^{\Ttran}(\posUE-\pos_{\text{RIS}})]_{1}),\label{eq:HWAODaz}\\
\phi_{\text{b}}^{\text{el}} & =\arccos\left([\vect{R}^{\Ttran}(\posUE-\pos_{\text{RIS}})]_{3}/\|\posUE-\pos_{\text{RIS}}\|\right),\label{eq:HWAODel}
\end{align}
where $\Delta_{\mathrm{clk}}\in \mathbb{R}$ is the user's clock bias, $c$ is the speed of light, $\vect{R}\in \mathbb{C}^{3 \times 3}$ is the rotation matrix defining the RIS's orientation (i.e., $\vect{R}^{\Ttran} \vect{z}$ maps $\vect{z}$ from the global to the local RIS coordinate system), and $[\vect{z}]_n$ is the $n$th entry of $\vect{z}$. Signal amplitudes may also be used in localization \cite{zhang2020towards} but this is not
explored here.

Without an RIS, estimating $\posUE$ from \eqref{eq:symbol-sampled-model-OFDM-Z10}
is impossible since the observation only yields $L_d-1$ TDOA measurements
$\{\tau_{d}^l-\tau_{d}^1: l=2,\ldots,L_d\}$, while there are $3L_d$ unknown geometric parameters: the user location
$\posUE$ and the locations of the $L_d-1$ scatter points $\pos_{\text{SP},l}$ (after removal of the clock bias).
However, we will show that adding a single RIS to the setup 
is sufficient to make the problem identifiable in terms of $\posUE$, though
not $\pos_{\text{SP},l}$. In particular, we will see the RIS acts as an additional synchronized BS with a phased array. 
We will describe the localization subproblems in detail:
design (offline RIS placement, online RIS configuration), channel
parameter estimation (determining $\{\tau_{\text{RIS}}, \vect{\phi}_{\text{b}}, \tau_{d}^l:l=1,\ldots,L_d\}$),
localization and synchronization (determining $\posUE$ and $\Delta_{\mathrm{clk}}$), and
sensing (determining $\{\pos_{\text{SP},l}: l=2,\ldots,L_d\}$).

\subsubsection*{RIS configuration encoding}
In localization, propagation paths with similar geometric parameters
(angles or delays) will not be resolved when two conditions are met:
(i) the delays and angles are similar; and (ii) they are correlated. Non-resolved
paths can lead to large biases in the estimates of
angles and delays. Making the RIS configuration $\overline{\vect{\omega}}_{\vect{\theta}}$ time-varying provides new dimensions to make paths resolvable. 
This can be achieved as follows over $T$ transmission blocks. 
We use a number $\beam\ll T$ RIS configurations  $\overline{\vect{\omega}}_{\vect{\theta}_1},\ldots,\overline{\vect{\omega}}_{\vect{\theta}_{\beam}}$
and associate a unique code (e.g., a column from a DFT matrix) $\vect{c}=[c_1,\ldots,c_{T/\beam}]^{\Ttran}\in\mathbb{C}^{T/\beam}$,
$|c_{n}|=1$ with a temporal balance property $\vect{c}^{\Ttran}\vect{1}=0$. 
The actual RIS configuration is
$c_t \overline{\vect{\omega}}_{\vect{\theta}_{\beamind}} $.
The switching is done in silent intervals between OFDM blocks to avoid modulating the reflected signals to other bands.
The received signal when using the $\beamind$th configuration is
\begin{equation}
\bar{\vect{z}}^{(\beamind)}[t]=\sum_{l=1}^{L_d} g^l \vect{d} (\tau_d^{l})\odot\bar{\vect{x}}+c_t g_{\text{RIS}} \vect{b}^{\Ttran}(\vect{\phi}_{\text{b}})\overline{\vect{\omega}}_{\vect{\theta}_{\beamind}} \left(\vect{d}(\tau_{\text{RIS}})\odot\bar{\vect{x}}\right)+\vect{w}^{(\beamind)}[t], \label{eq:HWcodedObservation} \quad t=1,\ldots,T/\beam
\end{equation}
where we use the same range of time indices for all configurations and separate them using the index $\beamind$.
The observations are grouped as $\vect{Z}^{(\beamind)}=[\bar{\vect{z}}^{(\beamind)}[1],\ldots,\bar{\vect{z}}^{(\beamind)}[T/\beam]]$,
from which we can compute an observation of the  uncontrollable
channel as $\hat{\vect{z}}^{(0)}=\sum_{\beamind=1}^{\beam}\vect{Z}^{(\beamind)}\vect{1}$
(with processing gain $T$) and of the $\beamind$th configuration of the controllable channel as $\hat{\vect{z}}^{(\beamind)}=\vect{Z}^{(\beamind)}\vect{c}^{*}$
(with processing gain $T/\beam$). This principle significantly
reduces complexity and storage at the RIS,
and is easy to generalize to a multi-RIS setup.

\subsection{RIS design for localization and sensing}

We want to  design an RIS-enabled localization system in a deployment
region $\mathcal{R} \subset \mathbb{R}^3$. We will rely on Fisher information theory (as developed for wideband localization in \cite{shen2010fundamental}, which we use as a basis in this article) for both offline and online design. 
We denote the unknown channel parameters as 
\begin{align}
\vect{\zeta}=[\tau_d^1,{\tau}_{\text{RIS}},\vect{\phi}_{b}^{\Ttran},\vect{\tau}_{l>1}^{\Ttran},\vect{g}^{\Ttran},g_{\text{RIS}}]^{\Ttran}.    
\end{align}
where $\vect{\tau}_{l>1} = [\tau_d^{2}, \tau_d^{3},\ldots, \tau_d^{L_d}]^{\Ttran}$ and $\vect{g} = [g^{1},g^{2},\ldots, g^{L_d}]^{\Ttran}$.
The design parameter vector $\vect{\sigma}$ accounts for the placement and configuration of the RIS and is selected from a 
 set $\mathcal{S}$.
The Fisher information matrix (FIM) can then be defined as 
\begin{equation}
\vect{J}(\vect{\vect{\zeta}}|\vect{\sigma})=\frac{2}{N_0}\sum_{t=1}^{T}\Re\left\{ \left(\nabla_{\vect{\zeta}}\vect{\mu}[t]\right)^{\Htran}\nabla_{\vect{\zeta}}\vect{\mu}[t]\right\} ,\label{eq:HWFIMdef}
\end{equation}
where $\vect{\mu}[t]=\bar{\vect{z}}[t]-\bar{\vect{w}}[t]$
is the noise-free observation, $\nabla_{\vect{\zeta}}\vect{\mu}[t] \in \mathbb{C}^{K\times(3L_d+5)}$ denotes the gradient, and $\Re$ returns the real part of its argument. The FIM satisfies the fundamental Fisher information
inequality $\vect{J}^{-1}(\vect{\vect{\zeta}}|\vect{\sigma})\preceq\mathbb{E}\{ (\vect{\vect{\zeta}}-\hat{\vect{\vect{\zeta}}})(\vect{\vect{\zeta}}-\hat{\vect{\vect{\zeta}}})^{\Ttran}\} $ (in the positive semidefinite sense),
under certain technical conditions, for any unbiased estimator $\hat{\vect{\vect{\zeta}}}$
of the channel parameters. We define a corresponding parameter vector in the location domain $\tilde{\vect{\zeta}}=[\posUE,\Delta_{\mathrm{clk}},\vect{\tau}_{l>1}^{\Ttran},\vect{g}^{\Ttran},g_{\text{RIS}}]^{\Ttran}$ 
and associated Jacobian $\vect{\Upsilon}=\nabla_{\tilde{\vect{\zeta}}}\vect{\zeta}$,
so that $\vect{J}(\tilde{\vect{\vect{\zeta}}}|\vect{\sigma})=\vect{\varUpsilon}^{\Ttran}\vect{J}(\vect{\vect{\zeta}}|\vect{\sigma})\vect{\varUpsilon}.$
From $\vect{J}(\tilde{\vect{\vect{\zeta}}}|\vect{\sigma})$, we can finally
compute the FIM of the user location using Schur's complement: we partition 
$\vect{J}(\tilde{\vect{\vect{\zeta}}}|\vect{\sigma})=\left[\vect{A}\,\vect{B};\vect{B}^{\Ttran}\vect{C}\right]$,
where $\vect{A}\in\mathbb{R}^{3\times3}$ so that $\vect{J}(\posUE|\vect{\sigma})=\vect{A}-\vect{B}\vect{C}^{-1}\vect{B}^{\Ttran}$. When
$\vect{J}(\posUE|\vect{\sigma})$ is invertible, we say that the
location is identifiable with $\vect{J}^{-1}(\posUE|\vect{\sigma})\preceq\mathbb{E}\left\{ (\posUE-\hat{\posUE})(\posUE-\hat{\posUE})^{\Ttran}\right\} $.
Since the FIM is a matrix, it is inconvenient as a design metric. However, the squared position error bound (SPEB) is a meaningful scalar metric (measured in $\text{m}^{2}$)
\begin{equation}
\text{SPEB}(\posUE|\vect{\sigma})=\text{trace}(\vect{J}^{-1}(\posUE|\vect{\sigma}))\le\mathbb{E}\{\|\posUE-\hat{\posUE}\|^{2}\}.
\end{equation}

\subsubsection{Offline design for optimized coverage}
A reasonable criterion for the offline
design phase is to provide uniform coverage or to maximize the fraction
of the deployment region with low SPEB. The latter criterion can be
expressed as 
\begin{align}
\underset{\vect{\sigma}\in\mathcal{S}}{\mathrm{maximize}}\,\,\, & \frac{1}{|\mathcal{R}|}\int_{\mathcal{R}}\mathbb{I}\left\{ \text{SPEB}(\posUE|\vect{\sigma})\le\varepsilon^{2}\right\} \text{d}\posUE \label{eq:HWofflineOpt}
\end{align}
where $\mathbb{I}$ is an indicator function, $|\mathcal{R}|$ is the size of the deployment region, and $\varepsilon$ is
a required accuracy (e.g., 1 m). Solving for $\vect{\sigma}$
leads to the optimal {placement of the RIS}. The problem (\ref{eq:HWofflineOpt})
can be solved by an exhaustive search over a finite set $\mathcal{S}$ ignoring
the uncontrollable channel, except for the LOS path, and using random RIS configurations $\overline{\vect{\omega}}_{\vect{\theta}_{\beamind}}$.

\subsubsection{Online design for optimized localization performance}

During the online design phase, we possibly have a priori information of
the location of the users and the origins of the uncontrollable channel. The
online problem to minimize worst-case localization performance is
then of the form
\begin{align}
\underset{\vect{\sigma}\in\mathcal{S}}{\mathrm{minimize}}\,\,\, & \max_{\posUE} \, \, \text{SPEB}(\posUE|\vect{\sigma})\label{eq:HWonlineOpt}
\end{align}
where $\vect{\sigma}$ includes the RIS configuration $\overline{\vect{\omega}}_{\vect{\theta}}$. The inner maximization $\max_{\posUE}$ is over
the high-probability region of the user location. This problem can
be rewritten as 
\begin{align}
\underset{\vect{\sigma}\in\mathcal{S},\vect{u}}{\mathrm{minimize}}\,\, & \vect{u}^{\Ttran}\vect{1} \label{eq:HWonlineOptCVX}\\
 \textrm{subject to}& \left[\begin{array}{cc}
\vect{J}(\tilde{\vect{\vect{\zeta}}}|\vect{\sigma}) & \vect{e}_{k}\\
\vect{e}^{\Ttran}_{k} & u_{k}
\end{array}\right]\succeq\vect{0},\,k=1,2,3,~\forall \tilde{\vect{\vect{\zeta}}},\notag
\end{align}
where $\vect{e}_{k}$ is a vector of zeros, except for a 1 in the $k$th entry, and constraints are added for each probable value of $\tilde{\vect{\vect{\zeta}}}$. 
The problem \eqref{eq:HWonlineOptCVX} is convex when the variable $\vect{\sigma}$ appears linearly in
$\vect{J}(\tilde{\vect{\vect{\zeta}}}|\vect{\sigma})$. 
The designs that minimize the SPEB are generally
different from those that maximize communication-centric metrics such as the capacity: though both have better performance at higher SNR, the localization accuracy depends also on the geometry and ability to separate, rather than align, signals from different paths. 

\subsection{Algorithms for estimation, localization, and sensing}

The algorithmic design depends on the underlying channel estimation method and the specific scenario. The algorithms can be Bayesian (i.e., providing a characterization of the distribution of the user and SP locations) or non-Bayesian (providing only a point estimate). A complete overview of such methods is out of the scope of this article. Instead, we  focus on single-antenna transmitters/receivers defined in \eqref{eq:symbol-sampled-model-OFDM-Z10}. 

\subsubsection{Algorithms for channel parameter estimation}

The controllable and uncontrollable channels
can be separated using the balanced code described in \eqref{eq:HWcodedObservation}. We obtain the following observation of the uncontrollable
channel:
\begin{equation}
\hat{\vect{z}}^{(0)}=\sum_{\beamind=1}^{\beam}\vect{Z}^{(\beamind)}\vect{1}=T\sum_{l=1}^{L_d} g^l \vect{d} (\tau_d^{l})\odot\bar{\vect{x}}+\sum_{t=1}^{T} \sum_{\beamind=1}^{\beam}\vect{w}^{(\beamind)}[t].\label{eq:HWpassivemultipath}
\end{equation}
Similarly, for the $\beamind$th configuration of the controllable channel ($\beamind\in\{1,\ldots,\beam\}$), we observe:
\begin{align}
\hat{\vect{z}}^{(\beamind)}=\vect{Z}^{(\beamind)}\vect{c}^{*} & =\frac{T}{\beam} g_{\text{RIS}} \vect{b}^{\Ttran}(\vect{\phi}_{b}) \overline{\vect{\omega}}_{\vect{\theta}_{\beamind}} \left(\vect{d}(\tau_{\text{RIS}})\odot\bar{\vect{x}}\right)+\sum_{t=1}^{T} \vect{w}^{(\beamind)} [t] c^*_t.\label{eq:HWactivemultipath}
\end{align}
Estimation of the uncontrollable channel can be performed using any standard
 channel estimation technique, e.g., compressive sensing \cite{venugopal2017channel}.
Estimation of the controllable channel involves only a single path (note
that multi-bounce reflections are very weak and hard to detect) and
can be performed using techniques from multi-antenna channel estimation
\cite{fascista2019millimeter}.

\subsubsection{Algorithms for localization and sensing}

After channel estimation, the interface towards localization and sensing
is via the estimated geometric channel parameters, say $\{\hat{\tau}_{d}^{l}:l=1,\ldots,L_d\}$,
$\hat{\tau}_{\text{RIS}}$, and $\hat{\vect{\phi}}_{b}$.
In the absence of a priori information, localization and sensing is
usually performed by determining an initial guess, based on the geometric
relations, followed by a refinement based on the likelihood function, which itself depends on the underlying channel estimation method.  
\begin{itemize}
\item \emph{Localization:} Assuming the LOS delay can be identified (e.g.,
from the large path power), \eqref{eq:HWtau0} and \eqref{eq:HWtaur}
leads a TDOA measurement, which defines a hyperbola with respect to the user position $\posUE$:
\begin{equation}
(\hat{\tau}_{\text{RIS}}-\hat{\tau}_{d}^{1})c\approx \|\posUE-\pos_{\text{RIS}}\|+\|\pos_{\text{BS}}-\pos_{\text{RIS}}\|-\|\posUE-\pos_{\text{BS}}\|,\label{eq:HWhyperbola}
\end{equation}
while measurements of the AOD in (\ref{eq:HWAODaz})--(\ref{eq:HWAODel}) 
determine a line from the RIS with direction 
\begin{equation}
\vect{k}(\hat{\vect{\phi}}_{b})\approx \frac{\vect{R}^{\Ttran}(\posUE-\pos_{\text{RIS}})}{\|\vect{R}^{\Ttran}(\posUE-\pos_{\text{RIS}})\|}.\label{eq:HWbearingline}
\end{equation}
The intersection of the hyperbola with the line determines
the user location, say $\hat{\posUE}$. An example will be provided later in Fig.~\ref{fig:localization_example}. Substituting $\hat{\posUE}$ back into
(\ref{eq:HWtau0}) provides us with an estimate of the clock bias,
say $\hat{\Delta}_{\mathrm{clk}}$. We note that in the presence of two RISs, the delay measurement
is not even needed, opening a path for accurate localization over
narrowband channels. These estimates can be refined with gradient
descent on the likelihood function. 
\item \emph{Sensing:} After the user location is determined, the sources
of uncontrollable channel is constrained by
\begin{equation}
(\hat{\tau}_{d}^{l}-\hat{\tau}_{d}^{1})c+\|\hat{\posUE}-\pos_{\text{BS}}\|\approx \|\hat{\posUE}-\pos_{\text{SP},l}\|+\|\pos_{\text{BS}}-\pos_{\text{SP},l}\|, \quad l>1.\label{eq:HWellipse}
\end{equation}
The right-hand side can be interpreted as a time-sum-of-arrival (TSOA),
which determines an ellipse with the BS and estimated user location as focal points.
 Since the controllable channel from the RIS is not dependent
on the uncontrollable channel, the RIS does not directly improve sensing,
but rather indirectly through better localization accuracy. Note that when there
is detectable multi-bounce multipath (BS to RIS to SP to the user), then
the AOD from the RIS to the SP can also be inferred, allowing unique
localization of each SP. 
\end{itemize}

\vspace{-4mm}

\subsection{Indoor localization example}

\begin{figure}
\begin{center} \vspace{-4mm}
	\begin{overpic}[width=\columnwidth,tics=10]{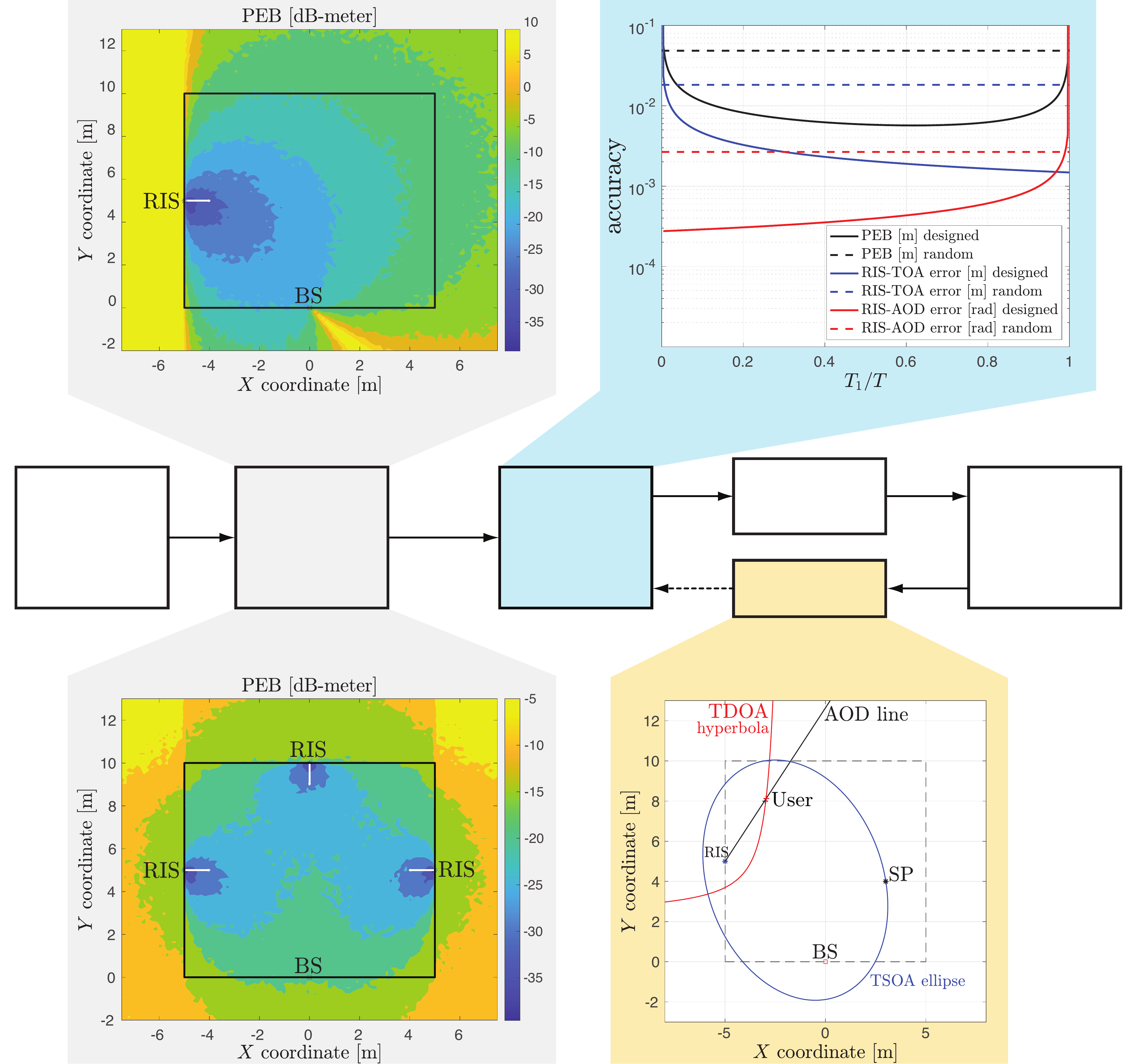}
		\put(4,47){Problem}
		\put(5.5,44.5){setup}
		\put(24,47){Offline}
		\put(24.5,44.5){phase}
		\put(48,47){Online}
		\put(48.5,44.5){phase}
		\put(68.3,50.8){Signal}
		\put(65.5,48.3){transmission}
		\put(65.5,41.3){Localization}
		\put(88,48.5){Channel}
		\put(87.5,46){parameter}
		\put(87.3,43.5){estimation}
		\put(26,55){(a)}
		\put(26,37){(b)}
		\put(55,55){(c)}
		\put(70,37){(d)}
\end{overpic} \vspace{-4mm}
\end{center} 
\caption{\label{fig:localization_example} The localization (and sensing) problem is solved in a sequence of steps, starting from the problem setup, then proceeding with the offline phase, online phase, and then the physical transmission, estimation, and localization. The online phase and localization can interact. (a) and (b) exemplifies the PEB (in dB-meter) over space for one or three RIS. The white lines show the normal to the RIS surface. (c) exemplifies the online design for a specific location, as a function of the fraction $T_1/T$ of configurations that maximizes the SNR over the total number of configurations. (d) exemplifies the localization output based on the LOS and controllable channel. The SP can be constrained to be on an ellipse.}   
\end{figure}

We will now exemplify the localization in a 2D scenario, where the elevation angle $\phi_{b}^{\text{el}}$ 
is removed from the set of unknown parameters to simplify the exposition. The methodology is summarized as a block diagram in Fig.~\ref{fig:localization_example} and we will describe the main blocks.
Following the scenario from Fig.~\ref{fig:Localization-and-sensing}, we
consider a $10\,\text{m}\times10\,\text{m}$ indoor environment with the
BS in the middle of a wall at location $\pos_{\text{BS}}=[0,0]^{\Ttran}$.
An RIS can be placed in the center of each of the three remaining walls.
The BS has an antenna that is omnidirectional in the azimuth plane and operates at a carrier frequency of $f_c=28$ GHz with 400
MHz bandwidth using $K=3000$ subcarriers and a transmission power of 20 dBm. The RIS consists of $N=64$ elements deployed along a line with $\lambda/5$
spacing (e.g., the total size is about 14 cm) and unity per-element gain  $G(\phi^{\text{az}})=1$ for $|\phi^{\text{az}}|\le\pi/2$ and $G(\phi^{\text{az}})=0$ elsewhere. The noise power spectral
density is $N_0=-174\,\text{dBm}/\text{Hz}$. We use $\beam=8$
RIS configurations and $T=256$ transmission blocks. The pilot symbols have constant modulus. We generate  \vspace{-2mm}
\begin{align}
g^{1} & =\frac{\lambda}{4\pi\|\posUE-\pos_{\text{BS}}\|}e^{\imagunit \psi^{1}},\\
g^{l} & =\frac{\lambda\sqrt{\sigma_{\text{RCS}}}}{(4\pi)^{3/2}}\frac{1}{\|\posUE-\pos_{\text{SP},l}\|}\frac{1}{\|\pos_{\text{SP},l}-\pos_{\text{BS}}\|}e^{\imagunit \psi^{l}}, \quad l>1 \\
g_{\text{RIS}} & =\sqrt{G(\phi_{a}^{\text{az}})G(\phi_{b}^{\text{az}})}\frac{(\lambda/5)^2}{4\pi}\frac{1}{\|\pos_{\text{RIS}}-\pos_{\text{BS}}\|}\frac{1}{\|\pos_{\text{RIS}}-\posUE\|}e^{\imagunit \psi_{\text{RIS}}}
\end{align}
where $\psi^{l},\psi_{\text{RIS}}$ are independently
and uniformly distributed in $[0,2\pi)$, while $\sigma_{\text{RCS}}$ is the radar
cross section (RCS) of the SP, expressed in $\text{m}^{2}$.

\subsubsection{FIM analysis}

It is instructive to investigate $\vect{J}(\posUE|\vect{\sigma})$
deeper for the case without uncontrollable multipath (only LOS) and a single RIS.
Using RIS configurations with temporal balance and balanced power allocation across subcarriers  
and an RIS phase reference in the center of the RIS, the FIM of the geometric parameters
$[\tau_{d}^1,\tau_{\text{RIS}},\phi_{b}^{\text{az}}]^{\Ttran}$ 
is
a diagonal matrix with entries (see \cite[Eq.~(16)--(17)]{shen2010fundamental} and \cite[Eq.~(4)]{Garcia18}) \vspace{-5mm}
\begin{align}
J(\tau_{d}^1)  =\frac{2|g^{1}|^{2}}{N_0}TB_{\text{eff}}^{2}\label{eq:HWFIMtauLOS}
\end{align}
\begin{align}
J(\tau_{\text{RIS}})  =\frac{2|g_{\text{RIS}}|^{2}\|\bar{\vect{x}}\|^{2}}{N_0}\frac{T}{\beam}B_{\text{eff}}^{2}\sum_{\beamind=1}^{\beam}\left|\vect{b}^{\Ttran}(\phi_{b}^{\text{az}})\overline{\vect{\omega}}_{\vect{\theta}_{\beamind}}\right|^{2}\label{eq:HWFIMtauRIS}
\end{align}
\begin{align}
J(\phi_{b}^{\text{az}})  =\frac{2|g_{\text{RIS}}|^{2}\|\bar{\vect{x}}\|^{2}}{N_0}\frac{T}{\beam}\bigg(\sum_{\beamind=1}^{\beam}\left|\dot{\vect{b}}^{\Ttran}(\phi_{b}^{\text{az}})\overline{\vect{\omega}}_{\vect{\theta}_{\beamind}}\right|^{2}-\frac{\left|\sum_{\beamind=1}^{\beam}\left(\vect{b}^{\Ttran}(\phi_{b}^{\text{az}})\overline{\vect{\omega}}_{\vect{\theta}_{\beamind}}\right)\left(\dot{\vect{b}}^{\Ttran}(\phi_{b}^{\text{az}})\overline{\vect{\omega}}_{\vect{\theta}_{\beamind}}\right)^{*}\right|^{2}}{\sum_{\beamind=1}^{\beam}\left|\vect{b}^{\Ttran}(\phi_{b}^{\text{az}})\overline{\vect{\omega}}_{\vect{\theta}_{\beamind}}\right|^{2}}\bigg)\label{eq:HWFIMAOD}
\end{align}
where $B_{\text{eff}}^{2}=\sum_{k=1}^{K} (2\pi k\Delta_{f})^{2}\left|\bar{x}_k\right|^{2}$ and $\dot{\vect{b}}(\phi_{b}^{\text{az}})$ denotes the derivative of ${\vect{b}}(\phi_{b}^{\text{az}})$ (defined in \eqref{eq:vector_b}) with respect to $\phi_{b}^{\text{az}}$. 
Based on the Jacobian, it is can be verified that the FIM becomes
\begin{align}
\vect{J}(\posUE|\vect{\sigma}) & =\frac{1}{c^{2}}\frac{J(\tau_{d}^1)J(\tau_{\text{RIS}})}{J(\tau_{d}^1)+J(\tau_{\text{RIS}})}(\vect{u}_{\text{BS}}-\vect{u}_{\text{RIS}})(\vect{u}_{\text{BS}}-\vect{u}_{\text{RIS}})^{\Ttran}+J(\phi_{b}^{\text{az}})\frac{\vect{\Xi}\vect{u}_{\text{RIS}}\vect{u}_{\text{RIS}}^{\Ttran}\vect{\Xi}^{\Ttran}}{\|\pos_{\text{RIS}}-\posUE\|^{2}},\label{eq:HWFIMclosedForm}
\end{align}
where $\vect{\Xi}=[0,\,-1;+1,\,0]$ is a rotation matrix over
$\pi/2$, $\vect{u}_{\text{BS}}$ is a unit vector from the BS to the user and $\vect{u}_{\text{RIS}}$ a unit vector from the RIS to the user.  The expression \eqref{eq:HWFIMclosedForm} shows that with aid of the RIS, we obtain two fundamental directions of Fisher information: 1) $\vect{u}_{\text{BS}}-\vect{u}_{\text{RIS}}$, with intensity (as defined in \cite{shen2010fundamental}) that depends on the TOA accuracy of both the LOS and RIS path; and 2) $\vect{u}_{\text{RIS}}$, with an intensity reduced with the distance.
This FIM analysis provides contradictory design requirements: for optimal
TOA estimation we should maximize the SNR and set $\overline{\vect{\omega}}_{\vect{\theta}_{\beamind}}=\vect{b}^{*}(\phi_{b}^{\text{az}})$,
for all $\beamind$. This is equivalent to the solution found in \eqref{eq:optimized-delays-basic} that maximizes capacity, but leads to $J(\phi_{b}^{\text{az}})=0$, meaning that the AOD cannot be estimated. On the other hand, for optimal AOD estimation \eqref{eq:HWFIMAOD} indicates that the RIS configurations $\overline{\vect{\omega}}_{\vect{\theta}_{\beamind}}$ should be a combination
of 
$\vect{b}^{*}(\phi_{b}^{\text{az}})$ from \eqref{eq:vector_b} and its derivative $\dot{\vect{b}}(\phi_{b}^{\text{az}})$.
Hence, a natural compromise is to configure the RIS using  $\overline{\vect{\omega}}_{\vect{\theta}_{\beamind}}=\vect{b}^{*}(\phi_{b}^{\text{az}})$ for
a fraction of the available transmissions, and 
set $\overline{\vect{\omega}}_{\vect{\theta}_{\beamind}}\approx\dot{\vect{b}}^{*}(\phi_{b}^{\text{az}})$
for the remaining transmissions (which involves approximating the derivative beam to be generated
by the RIS so that $|[\overline{\vect{\omega}}_{\vect{\theta}_{\beamind}}]_{n}|=1$).
By optimizing the fraction, the two terms in (\ref{eq:HWFIMclosedForm})
can be balanced. The RIS essentially behaves like
an additional synchronized BS equipped with a phased array.

\subsubsection{Offline design}

We first consider five alternative designs: no RIS, an RIS on the left
wall, an RIS on the front wall (facing the
BS), an RIS on the right wall, or three RISs (one on each remaining wall, using orthogonal temporally balanced codes).
Random RIS phase configurations were assumed. 
Setting the required accuracy to $\varepsilon=0.1\,\text{m}$ in \eqref{eq:HWofflineOpt}, the fraction of locations that have sufficiently low PEB is 
$0$ (no RIS), $0.35$ (left RIS),
$0.45$ (facing RIS), $0.35$ (right RIS) and $0.99$ (three RIS). This shows that it is
better to put the RIS on the wall facing the RIS (despite larger propagation loss), and that using three RISs can provide uniform coverage in the deployment
region. To gain further insight, Fig.~\ref{fig:localization_example}(a,b) shows
a contour plot of $\text{PEB}(\vect{x|}\vect{\sigma})=\sqrt{\text{SPEB}(\vect{x|}\vect{\sigma})}$
for two of the configurations $\vect{\sigma}$ (RIS on the left wall
and three RISs). For visualization purposes, the PEB is expressed in dB-meter (i.e., $10 \log_{10}(\text{PEB})$), where $0\,\text{dB-meter}$ means 1 meter uncertainty, $-10\,\text{dB-meter}$ is 0.1 meter uncertainty, etc.
We see that when an RIS is placed on the left wall, low
PEB is achieved only very close to the RIS, especially in the lower
part of the room, closer to the BS. Along the line between the BS and
RIS, behind the BS, the PEB tends to infinity, since the unit vector from the BS to the user $\vect{u}_{\text{BS}}$ is parallel to the unit vector from the RIS to the user $\vect{u}_{\text{RIS}}$, leading to  $\vect{u}_{\text{BS}}-\vect{u}_{\text{RIS}}=\vect{0}$ in \eqref{eq:HWFIMclosedForm},
so the TDOA measurement brings no information. Behind the RIS, the
PEB is also infinite, due to the zero per-element gain $G(\phi^{\text{az}})$. On
the other hand, with three RISs, we can obtain acceptable performance throughout
the deployment region.

\subsubsection{Online design}

We now use a single RIS on the left wall (see Fig.~\ref{fig:localization_example}(a))
and a user at $\posUE=[-3,8]^{\Ttran}$. We aim to optimize
the RIS configuration for this location and consider the following
alternatives (which describe the set $\mathcal{S}$ of design variables):
(i) set the RIS configurations $\overline{\vect{\omega}}_{\vect{\theta}_{\beamind}}$ to be random, $\beamind=1,\ldots,8$, each configuration is used 32 times; (ii) in the $T=256$ transmissions, use $T_1$
times the direct beam configuration $\overline{\vect{\omega}}_{\vect{\theta}_{\beamind}}=\vect{b}^{*}(\phi_{b}^{\text{az}})$
and $T-T_1$ times the approximation of the derivative beam configuration $\overline{\vect{\omega}}_{\vect{\theta}_{\beamind}}\approx\dot{\vect{b}}^{*}(\phi_{b}^{\text{az}})$.
In Fig.~\ref{fig:localization_example}(c), we evaluate, as a function of $T_1/T$,
the $\text{PEB}(\posUE|\vect{\sigma})$, the error standard
deviation of $c \tau_{\text{RIS}}$, $c \sqrt{J^{-1}(\tau_{\text{RIS}})}$, and the error standard deviation of $\phi_{b}^{\text{az}}$, given by
$\sqrt{J^{-1}(\phi_{b}^{\text{az}})}$. We recall that $T_1=T$
is optimal in terms of SNR and TOA estimation accuracy. From a localization perspective, the best performance is achieved when $T_1/T\approx0.63$,
while for $T_1=0$ and $T_1=T$, the PEB diverges. This behavior
can be explained by inspecting the TOA and AOD: a large
$T_1$ leads to high SNR and maximizes $J(\tau_{\text{RIS}})$,
so the best TOA estimation from the RIS is achieved when only $\overline{\vect{\omega}}_{\vect{\theta}_{\beamind}}=\vect{b}^{*}(\phi_{b}^{\text{az}})$
is used. However, in that case, $J(\phi_{b}^{\text{az}})\to0$ so that
the measurement does not provide any information about the AOD from
the RIS. While not obvious from the figure, when only $\overline{\vect{\omega}}_{\vect{\theta}_{\beamind}}\approx\dot{\vect{b}}^{*}(\phi_{b}^{\text{az}})$
is sent, $J(\tau_{\text{RIS}})=J(\phi_{b}^{\text{az}})=0$, since $\vect{b}^{\Ttran}(\phi_{b}^{\text{az}})\dot{\vect{b}}^{*}(\phi_{b}^{\text{az}})=0$.
The random configurations (though providing information when there is no prior on the user location) lead to worse PEB than the optimally designed configurations.

\subsubsection{Localization and sensing}

As a final example, we again use a single RIS on the left wall, a user
at $\posUE=[-3,8]^{\Ttran}$, and an SP at location $\pos_{\text{SP}}=[3,4]^{\Ttran}$
with RCS of $\sigma_{\text{RCS}}=1\,\text{m}^{2}$. To estimate the delays of the uncontrollable
channel, we apply a DFT to (\ref{eq:HWpassivemultipath}) and determine
the peaks. These can then be converted to $\hat{\tau}_{d}^{1}$ and $\hat{\tau}_{d}^{2}$.
To estimate the TOA and AOD from the controllable channel, we use (\ref{eq:HWactivemultipath})
and perform a 2D search over $[\tau_{\text{RIS}},\phi_{b}^{\text{az}}]$
with a substituted estimate of the channel gain. This yields $\hat{\tau}_{\text{RIS}}$
and $\hat{\phi}_{b}$. Fig.~\ref{fig:localization_example}(d)
shows the locations of the BS, user, RIS, and SP, the TDOA hyperbola
from (\ref{eq:HWhyperbola}) as well as the AOD bearing line from
(\ref{eq:HWbearingline}). Their intersection is the estimated location
$\hat{\posUE}$. From this estimated location and the TOA of the
uncontrollable channel, we obtain a TSOA ellipse (\ref{eq:HWellipse}),
near which the SP must lie. Note that from a snapshot, the SP location
cannot be determined, but after sufficient movement of the user and appropriate data association, the
SP location can be uniquely identified.

\vspace{-3mm}

\subsection{Conclusions from RIS-aided localization}

An RIS can be seen as a synchronized multi-antenna
BS with a phased array that can aid the localization. Proper RIS placement  can provide
significant location coverage improvements. The RIS configuration can be tailored to
the specific user location, but is significantly different from the optimal configuration for communications. This is also reflected by the different simulation setups considered in this article:
many RIS elements are required in communications to improve the end-to-end SNR, while localization requires large bandwidths but the RIS can be small since it is primarily used to add new dimensions to resolve identifiability issues.  An increase in the number of RIS elements can enable the use of less bandwidth. 
While for communication, the RIS provides limited gains when the LOS path is present, for localization both paths provide useful and necessary information. 
Moreover, by encoding the RIS configurations
with a global code, the controllable and uncontrollable channels can be separated,
and multiple non-interfering RISs can be supported. The use of RIS
for sensing is mainly indirect, by the improved estimation of the
user location, if the uncontrollable and controllable channels do not interact.

\enlargethispage{6mm}

\section{Future model evolution and related signal processing challenges}

We next elaborate on a few fundamental phenomena that appear when having a large and dense RIS. By refining the models to capture these properties, there are opportunities to develop new signal processing algorithms that push the boundaries of how communications and localization are normally conceived.

\subsection{Scaling laws and near-field regime}

The benefit of classical beamforming from an antenna array is that the SNR grows linearly with the number of antennas $N$ \cite{Veen1998a}. 
When maximizing the narrowband capacity, we noticed that the SNR with an RIS instead grows as $N^2 \alpha \beta\gamma$, when the $N$ paths have the same propagation loss.
The quadratic SNR scaling does not mean that the setup in Fig.~\ref{fig:basic_example}(b) can achieve a higher SNR than in a case where the RIS is replaced by an equal-sized antenna array that is transmitting with equal power. In the latter case, the SNR would be proportional to $N\beta$. To understand the difference, we can factorize the SNR scaling achieved by the RIS as $( N\alpha \gamma) \cdot (N\beta)$. The first term accounts for the fraction of the transmitter's signal power that is reflected by the RIS, which is a very small number even when $N$ is large since $\alpha < -70$\,dB is typical.
Hence, the RIS cannot achieve a higher SNR than $N \beta$ but the difference reduces as $1/N$. When comparing an RIS with alternative technologies, the RIS must be physically larger to be competitive \cite{Bjornson2020a}.

SNR scaling behaviors are extensively studied in signal processing for communications and localization to understand the ultimate performance and obtain intuitive performance approximations for cases with large arrays.
Although the asymptotic regime where $N\to \infty$ is commonly studied, practical technologies have thus far operated far from the limit so it has been unimportant whether the underlying models are asymptotically accurate or not.
Since the law of conservation of energy dictates that we cannot receive more power than what was transmitted, the SNR must approach a finite upper limit as $N\to \infty$. The aforementioned SNR scaling was obtained under a far-field assumption: the propagation losses $\alpha_n \beta_n$ are equal for all $N$ RIS elements.
However, when the transmitter and/or receiver is at a distance from the RIS similar to the RIS's width/height, the geometry will make $\alpha_n \beta_n$ widely different between the elements \cite{BS20}. This scenario is unavoidable as $N \to \infty$ but also occurs in practice when the RIS is 1\,m wide and the user is at a similar distance.
The general SNR expressions presented in this tutorial remain valid and if propagation loss models that capture the radiative near-field properties are utilized, one can derive how the SNR converges to a finite upper limit \cite{BS20}.
More importantly, the near-field enables the RIS to focus signals not only in a particular direction but at a certain point in that direction, thereby making a flat RIS better than a flat mirror \cite{Bjornson2020a}. This property can also be utilized for improved localization \cite{elzanaty2020reconfigurable}.
Since classical array signal processing focuses on the far-field, there are great opportunities to develop new algorithms that exploit the unique near-field properties for improved communication and localization.

\enlargethispage{2mm}

\vspace{-4mm}

\subsection{Channel modeling and sparsity}

The system models in this article can be utilized for any LTI channel, but the channel modeling for RIS is yet in its infancy with limited experimental validation.
In a multipath environment, different parts of the RIS will observe different linear combinations of the impinging waves, leading to fading variations.
The wavelength limits the variability and, even in rich scattering, there will be correlation between the channel coefficients observed at RIS elements that are within a few wavelengths \cite{Bjornson2021a}.
This fundamental property has several impacts on the RIS operation. The resulting spatial sparsity can be exploited to simplify channel estimation. It also enables an RIS to reflect multiple signals to different locations simultaneously, thereby enabling communication or localization with multiple users. 
The modeling of how an RIS interacts with interference from other systems and electromagnetic noise remains open.
There are signal processing research challenges in both system modeling, algorithmic design, and optimization.

\enlargethispage{5mm}

\vspace{-3mm}

\subsection{Non-linear RIS operation}

This article focuses on applications where the RIS has a (piecewise) constant configuration, so it can be modeled as a linear filter. We also explained how LTV system theory can be used in the case where the RIS is tuned to mitigate the Doppler effect caused by mobility.
A different option is to vary the RIS configuration continuously during the transmission of a signal block to modulate the transmitted signal before it is reradiated \cite{Yuan2021a}.
This effectively creates a non-linear end-to-end channel where the received signal contains a wider range of frequencies than the transmitted signal. The key applications remain to be discovered but it is clear that signal processing provides the right tools for analysis and optimization.

\vspace{-3mm}

\subsection{Mutual coupling}

A model assumption that was made already in Fig.~\ref{fig:basic_example}(b) is that the $N$ elements act as separate filters that each take a single input. However, when the RIS elements are closely spaced, it is hard to fully isolate them on the substrate material. This leads to mutual coupling where the impedance of one element is connected with the impedances of the neighboring elements. Hence, if the curves in Fig.~\ref{fig:basic_example} exemplify how an RIS element behaves in isolation, in reality, the frequency response will also depend on the configuration of the neighboring elements.
The mutual impedance is dependent on the physical properties of the elements and can be determined through lengthy full-wave simulations, such as the method of moments, that must be carried out for each configuration. 
Such an analysis has previously been done for antenna arrays.
The special case of canonical minimum-scattering (CMS) antennas \cite{wasylkiwskyj_theory_1970} allows expressing the mutual impedance as a closed-form function of the distance and orientations of two antennas \cite{gradoni_end--end_2020,williams_multiuser_2020}.

The CMS approach does however not capture the desired operation of the RIS as CMS antennas do not allow full $2\pi$ phase control \cite{qian_mutual_2020}. 
In contrast, an RIS made from patch or slot antennas effectively decouples the amplitude and phase of the reflected wave, allowing full $2\pi$ phase control.
Without proper modeling of the mutual impedance, the trade-off between complexity and performance as the RIS is densified cannot be evaluated. As such, different modeling techniques that do not rely on the CMS assumption have to be developed to capture the behavior of RISs with closely spaced elements.

For conventional antenna arrays, mutual coupling carries drawbacks such as scan blindness and ohmic losses. Scan blindness is when the wave is fully reflected and might be a desirable effect for an RIS. The high ohmic losses of superdirectivity could potentially be utilized to absorb interference as heat. As such, these effects present themselves as new opportunities that should be reevaluated in the RIS context.

A methodology based on circuit theory can be utilized to develop discrete-time RIS system models that capture mutual coupling, but this research is yet in its infancy. Mutual coupling will  have an impact on the algorithmic design as well as the communication/localization performance. If accurate models are hard to develop, machine learning methods might be useful to address the problem of system identification.

\vspace{-4mm}

\section{Summary}

This article has provided a tutorial of the basic system modeling of wireless signaling that involves RIS. This emerging technology can be utilized to increase the capacity of communication systems and the accuracy of localization and sensing systems. While the same models underpin both applications, the preferred embodiments differ in terms of bandwidth requirements, RIS dimensions, and optimal configuration. The basic algorithms and properties have been described in this article, but there is a goldmine of open signal processing problems, for example, related to refined models capturing the relevant electromagnetic properties, experimental validation, and more realistic applications.
Since RIS technology is often mentioned in 6G research, \emph{now is the right time} to explore these open problems.

\vspace{-3mm}

\enlargethispage{5mm}

\section*{Acknowledgments}

We would like to thank Gonzalo Seco-Granados, Kamran Keykhosravi, \"Ozlem Tugfe Demir, and Robin J. Williams for their comments and feedback during the writing. This work has been partially supported by H2020 RISE-6G project, under grant 101017011, the German Research Foundation (DFG) under Germany's Excellence Strategy (EXC 2077 at University of Bremen, University Allowance), the Italian Ministry of Education and Research in the framework of the CrossLab Project, and  the FFL18-0277 grant from the Swedish Foundation for Strategic Research.

\section*{Authors}

{\textbf{Emil Bj\"ornson} (emilbjo@kth.se) received his Ph.D. degree from the KTH Royal Institute of Technology, Sweden, in 2011. He is currently a Professor at the KTH Royal Institute of Technology. His research interests are MIMO and RIS-aided communications, radio resource allocation, and energy efficiency. He has received the 2018 IEEE Marconi Prize Paper Award, the 2019 EURASIP Early Career Award, the 2019 IEEE Fred W. Ellersick Prize, the 2020 Pierre-Simon Laplace Early Career Technical Achievement Award, the 2020 CTTC Early Achievement Award, and the 2021 IEEE ComSoc RCC Early Achievement Award. He is a Fellow of IEEE.}

{\textbf{Henk Wymeersch} (henkw@chalmers.se) received his Ph.D. degree from Ghent University, Belgium, in 2005. He is currently a Professor in Communication Systems at Chalmers University of Technology, Sweden, and Distinguished Research Associate with Eindhoven University of Technology, The Netherlands. His current research interests are in the convergence of communication, localization, and sensing. He was co-recipient of the Best Paper Award and Best Student Paper Award at the 2021 International Conference on Information Fusion (FUSION).}

{\textbf{Bho Matthiesen} (matthiesen@uni-bremen.de) received his Ph.D. degree from TU Dresden, Germany, in 2019. He is currently a research group leader at the U Bremen Excellence Chair of Petar Popovski in the Department of Communications Engineering, University of Bremen, Germany. His research interests are in communication theory, wireless communications, and optimization theory. He is an Exemplary Reviewer 2020 of the IEEE Wireless Communications Letters, was an invited speaker at the 2nd 6G Wireless Summit 2020, and served as a publication chair for the International Symposium on Wireless Communication Systems (ISWCS) 2021.}

{\textbf{Petar Popovski} (petarp@es.aau.dk) is a Professor Aalborg University, where he heads the section on Connectivity. He received Dipl.-Ing. (1997)/Mag.-Ing. (2000) in communication engineering from Sts. Cyril and Methodius University in Skopje and Ph. D. from Aalborg University (2004). He is a Fellow of IEEE, holder of an ERC Consolidator Grant (2015-2020), Villum Investigator, and a Member at Large on the Board of Governors in IEEE Communication Society. His research interests are in wireless communications/networks and communication theory. He authored the book “Wireless Connectivity: An Intuitive and Fundamental Guide”, published by Wiley in 2020.}

{\textbf{Luca Sanguinetti} (luca.sanguinetti@unipi.it) received the Laurea degree (cum laude) in telecommunications engineer and the Ph.D. degree in information engineering from the University of Pisa, Italy, in 2002 and 2005, respectively. He is currently an Associate Professor at the University of Pisa, Italy. His main research interests span the areas of wireless communications and signal processing for communications, with particular emphasis on multiple antenna technologies. He coauthored two textbooks: \emph{Massive MIMO Networks: Spectral, Energy, and Hardware Efficiency} (2017) and \emph{Foundations of User-Centric Cell-Free Massive MIMO} (2020).  He received the 2018 Marconi Prize Paper Award in Wireless Communications.}

{\textbf{Elisabeth De Carvalho} (edc@es.aau.dk) received the Ph.D. degree in Electrical Engineering from Telecom ParisTech, France. She is now a Professor at Aalborg University, Denmark. She has co-authored the book A Practical Guide to MIMO Radio Channel. Her main expertise is in the field of signal processing with emphasis on MIMO communications. She is a member of IEEE Signal Processing Society, the SPCOM technical committee and vice chair of the IEEE ComSoc ETI on Machine Learning for Communications. She is an associate editor of IEEE Transactions of Wireless Communications. She is the coordinator of the European Union H2020 ITN WindMill.}

\bibliographystyle{IEEEtran}
\bibliography{IEEEabrv,refs}

\end{document}